\documentclass[twocolumn]{aastex63}
\usepackage{graphicx}
\usepackage{dcolumn}
\usepackage{bm}
\usepackage{amsfonts}
\usepackage{color}
\usepackage{newtxtext,amsmath}
\usepackage{appendix}
\usepackage{subfigure}

\usepackage{lineno}

\begin{document}

\title{Core-collapse supernova simulations and the formation of neutron stars, hybrid stars, and black holes}

\correspondingauthor{Takami Kuroda}
\email{takami.kuroda@aei.mpg.de}

\author{Takami Kuroda}
\affiliation{Max-Planck-Institut f{\"u}r Gravitationsphysik, Am M{\"u}hlenberg 1, D-14476 Potsdam-Golm, Germany}

\author{Tobias Fischer}
\affiliation{Institute of Theoretical Physics, University of Wroclaw, Plac Maksa Borna 9, 50-204 Wroclaw, Poland}

\author{Tomoya Takiwaki}
\affiliation{Devision of Science, National Astronomical Observatory of Japan, 2-21-1, Osawa, Mitaka, Tokyo 181-8588, Japan}

\author{Kei Kotake}
\affiliation{Department of Applied Physics and Research Institute of Stellar Explosive Phenomena, Fukuoka University, Jonan-ku, Fukuoka City, Fukuoka 814-0180, Japan}

\shorttitle{The formation of NSs, hybrid stars, and BHs}
\shortauthors{Kuroda et al.}

\begin{abstract}
We investigate observable signatures of a first-order quantum chromodynamics (QCD) phase transition in the context of core collapse supernovae. To this end, we conduct axially symmetric numerical relativity simulations with multi-energy neutrino transport, using a hadron-quark hybrid equation of state (EOS).  We consider four non-rotating progenitor models, whose masses range from $9.6$ to $70$\,M$_\odot$. We find that the two less massive progenitor stars (9.6 and 11.2\,M$_\odot$) show a successful explosion, which is driven by the neutrino heating. They do not undergo the QCD phase transition and leave behind a neutron star (NS). As for the more massive progenitor stars (50 and 70 M$_\odot$), the proto-neutron star (PNS) core enters the phase transition region and experiences the second collapse. Because of a sudden stiffening of the EOS entering to the pure quark matter regime, a strong shock wave is formed and blows off the PNS envelope in the 50 M$_\odot$ model. Consequently the remnant becomes a quark core surrounded by hadronic matters, leading to the formation of the hybrid star. However for the 70 M$_\odot$ model, the shock wave cannot overcome the continuous mass accretion and it readily becomes a black hole. We find that the neutrino and gravitational wave (GW) signals from supernova explosions driven by the hadron-quark phase transition are detectable for the present generation of neutrino and GW detectors. Furthermore, the analysis of the GW detector response reveals unique kHz signatures, which will allow us to distinguish this class of supernova explosions from failed and neutrino-driven explosions.
\end{abstract}


\section{Introduction}
\label{sec:Introduction}
The final fate of massive stars is the moment of birth of diverse compact stars.
Massive stars heavier than $\sim8\,{\rm M}_\odot$ undergo gravitational collapse of their iron core once their core mass reaches the Chandrasekhar limit.
Thereafter part of or whole the star collapses and becomes a compact remnant star: a neutron star (NS) or a black hole (BH) or an {\it exotic} star.
The remnant type is determined mostly from the progenitor mass \citep{Heger03}.
Generally speaking, more massive stars tend to leave behind BHs and conversely less massive stars have more chances of leaving NSs \citep{Fryer99}.

The gravitational collapse releases huge amount of gravitational binding energy, part of which is deposited into the stellar envelope.
Depending on the progenitor star, the energy deposition can trigger the stellar explosion called core-collapse supernova (CCSN).
The explosion dynamics regulates how much mass is blown off, or equivalently how much mass accretes on the remnant star.
Therefore, the detailed explosion dynamics including the success or failure of the explosion are the keys to understand the diversity of remnant compact stars.

Two major explosion mechanisms are currently proposed.
The first one relies on the complex interplay between neutrinos and stellar mantle, called neutrino-driven explosion (NDE).
It is considered to account for the vast majority of observed CCSNe, showing an explosion energy of $\sim10^{51}$\,erg \citep[for recent reviews, see][]{Janka16,Radice18,BMulelr20_review}.
The other is called magnetorotational explosion (MRE) that takes place if the progenitor star rotates sufficiently fast and is also magnetized \citep[see,][for recent three-dimensional studies]{Moesta18,Obergaulinger21,KurodaT21,Bugli21}.
Due to its efficient energy conversion mechanism, the MRE mechanism may explain some exceptional explosive events called hypernovae \citep{iwamoto98,Nomoto06}, which exhibit one order of magnitude larger explosion energy than the canonical events.
Roughly speaking, the NDE and MRE are currently presumed to explain CCSNe of less and more massive stars, respectively.

Recently another explosion mechanism gains more attention in the context of both the explosion dynamics and formation process of neutron stars composed of exotic matter at their interiors \citep[c.f.][]{Fischer18,Zha20}.
When the energy density becomes sufficiently high then nuclear, in general hadronic, matter undergoes a phase transition to the deconfined quark-gluon plasma. Therefore, the solution of quantum chromodynamic (QCD)---the theory for strong interactions with quarks and gluons as the degrees of freedom---predicts a smooth cross-over transition at a pseudocritical temperature of in the range of 150--160\,MeV~\cite{Bazavov:2014,Borsanyi:2014,Bazavov:2019}, however, only at vanishing baryon density and hence cannot be applied to astrophysical applications of compact stellar objects such as neutron stars and supernovae. At high density phenomenological quark matter models have long been employed in astrophysical studies, based on the common two-phase approach with separate hadronic and quark matter phases. Such phase transition construction results not only in the transition to the third family of compact stars, known as $hybrid$ stars, as an extension to the neutron stars second family, but also in the yet incompletely understood question about the nature of the QCD phase transition at high baryon density \citep[c.f.][and references therein]{Annala20}.  
It is likely that low- and intermediate-mass NSs, having masses of $\sim$1.4\,M$_\odot$, can be explained solely by hadronic equations of state (EOS).
More massive NSs with $\sim$2\,M$_\odot$, whose presence is observationally confirmed at high accuracy~\citep[c.f.][]{Demorest10,Antoniadis13,Cromartie:2020NatAs}, feature the highest baryon densities encountered in the universe. It is a valid question to ask if a QCD transition occurs at the densities encountered at the interior of such massive pulsars \citep[c.f.][and references therein]{Kurkela:2014vha}. Furthermore, if confirmed, this is likely to affect the understanding of the core-collapse supernova phenomenology.

This is the origin of the third explosion mechanism, which utilizes the difference in the gravitational binding energy between NSs and hybrid stars.
As a consequence, once the central energy density of the PNS enters the phase transition region, the PNS core migrates to a new state, liberating considerable amount of gravitational potential energy. 
The latter is directly related to the latent heat, as a consequence of the phase transition construction within the two-phase approach.
\citep[for a recent review, see][]{Baym18_Quark_review}. Assuming a sufficiently strong first order transition, previous CCSN simulations showed a possible explosion scenario \citep{Sagert09,Nakazato13,Fischer18,Zha20}, which relies on the energy liberation at the moment of the phase transition.
Under such conditions, the PNS core favors a dynamical collapse, rather than a smooth transition, into the hybrid star branch.
When the infalling matter collides with a solid surface of the quark core, the dynamical collapse can be halted depending on the PNS mass.
Then it leads to a formation of a strong shock wave, which may blow off the outer envelope.
This is the mechanism of the explosion driven by the QCD phase transition (hereafter, phase transition driven explosion, PDE).

It is intriguing to explore the progenitor mass dependence on the PDE dynamics and also on the final remnant property.
One can naively expect that progenitors on the less massive side (say $\lesssim20$\,M$_\odot$) may explode as CCSNe \citep[e.g.,][]{BMuller13,takiwaki16,Vartanyan19b}, probably by the standard NDE mechanism, and leave NSs. 
This is because, for such low mass stars, the PDE may not be relevant, as the PNS interior unlikely becomes so dense enough to reach the QCD phase transition before the NDE.
While the more massive stars ($\gtrsim70$\,M$_\odot$) tend do become failed CCSNe and leave BHs \citep{KurodaT18,Shibagaki21}.
Although the progenitor mass threshold, above which they are expected to end up with BH formations, is yet to be understood, some intermediate-mass stars, e.g., $\sim40-50$\,M$_\odot$ at their zero-age main sequence (ZAMS), are reported to form BHs in the standard neutrino heating mechanism \citep[irrespective of the spatial dimension:][]{Liebendorfer04,Sumiyoshi07,Chan&Muller18,Pan18}.
However, with the aid of the QCD phase transition, these intermediate-mass stars could turn into successful CCSNe with the remnant being hybrid stars \citep{Fischer18}.
Progenitor mass dependence of phase transition is studied by \cite{Fischer20,Zha21}, though only in spherically symmetric models.

It is noteworthy that, contrary to the NDE case, the PDE mechanism works even in the spherically symmetric model.
There is a consensus that the successful CCSN explosion driven by neutrino heating cannot be achieved under the spherical symmetry (1D) \citep{Liebendorfer01,Sumiyoshi05}, except for a few less massive star models \citep[with $\lesssim10$\,M$_\odot$,][]{kitaura}.
This is because the inefficient neutrino heating in the absence of lateral hydrodynamic motions in 1D.
The mechanism of PDE, on the contrary, does not rely on the neutrino heating, but mostly on the strongness of hydrodynamic bounce of quark core, i.e. on the QCD phase transition.
Therefore most of previous PDE studies are reported using 1D models \citep{Sagert09,Nakazato13,Fischer18,Fischer:2020} and the only multi dimensional (multi-D) model reported so far is by \cite{Zha20}.
However, their study focuses mainly on the impacts of phase transition on the gravitational wave (GW) emission using a particular progenitor star with $12$\,M$_\odot$, and the hydrodynamic evolution of multi-D PDE model is thus currently not well studied.

There are additional remarkable aspects of the PDE as follows: its relatively large explosion energy of a few times $10^{51}$\,erg \citep{Fischer18} and a possible mass window of its progenitor stars.
For the latter, the intermediate-mass stars with $\sim40$-$50$\,M$_\odot$ at ZAMS stage could be the potential candidates of the PDE as less massive stars ($\lesssim20$\,M$_\odot$) may explode solely by the neutrino heating mechanism, while too massive stars can directly collapse into a BH \citep{Heger03}.
According to the stellar evolution models, such intermediate-mass stars experience severe mass loss during their evolutionary stages \citep[e.g.,][]{Umeda08} and would produce a dense circumstellar medium (CSM).
From these facts, the PDE might be able to account for a subclass of CCSNe: the CSM-interacting superluminous SN (hereafter SLSN).
This is because the observed light curves of SLSNe, such as SN~2006gy \citep{Smith07,Moriya13} or SN~2008fz \citep{Drake10_sn2008fz}, can be well modeled by slightly larger explosion energy than normal type IIp SNe \citep{Hamuy03}, a few times $10^{51}$\,erg,  and a dense CSM, with which the SN explosion interacts.
We should, however, note that the origin of such superluminous transient events are still under debate.
While the above-mentioned interaction between type IIn SN ejecta and a dense CSM is indeed one plausible mechanism, other possible scenarios are also proposed to explain the light curves of SLSNe: such as by  pulsational pair instability supernovae \citep{Woosley2007} or type Ia SNe in CSM \citep{Jerkstrand20_sn2006gy_Ia}.


In this study, we perform 2D axially symmetric numerical simulations of core collapse of massive stars and explore the possible impacts of the phase transition on the PNS evolution and the remnant property.
We particularly focus on the PDE dynamics in multi-D model.
To this end, we select four massive progenitor stars ranging from $9.6$\,M$_\odot$ to $70$\,M$_\odot$ and use one particular EOS, which considers the first order phase transition from hadronic matter to quark-gluon plasma \citep[for details, see][]{Fischer18,Bastian:2021}.
The employed supernova model is based on numerical relativity and solves the general relativistic  neutrino radiation hydrodynamics equations together with the two-moment (M1) neutrino transport equations \citep[details can be found in][]{KurodaT16}.
We use up-to-date neutrino opacities following \citet{Kotake18}.
Our results clearly show a diversity of postbounce evolution for various progenitor masses.
We also discuss their multi-messenger signals: gravitational waves (GWs) and neutrino signals.

This paper is organized as follows.
Section~\ref{sec:Method and initial models} starts with a concise summary of our GR radiation-hydrodynamic (RHD) scheme and also describe the initial setup of the simulation.
The main results and detailed analysis of the effects of phase transition are presented in Section~\ref{sec:Results}.
In Section~\ref{sec:GW and neutrino emission}, we discuss multi-messenger signals obtained from computed models.
We summarize our results and conclude in Section~\ref{sec:Summary and Discussion}.
Note that cgs units are used throughout the paper.
Greek indices run from 0 to 3 and Latin indices from 1 to 3, except $\nu$ and $\varepsilon$ that denote neutrino species and energy, respectively.

\section{Method and initial models}
\label{sec:Method and initial models}
In our full GR RHD simulations, we solve the evolution equations of geometrical variables and the relativistic hydrodynamics together with the energy-dependent neutrino radiation transport.
The basic part in solving the evolution equations is the same as our previous full 3D studies in the Cartesian coordinates \citep{KurodaT16,KurodaT20}.
The only difference is that we employ the so-called Cartoon method \citep{Shibata00_Cartoon,Alcubierre01_Cartoon}, which allows us to evolve the axisymemtric geometrical variables in the 3D Cartesian coordinates.
On the other hand, we use the cylindrical coordinates for the RHD part, which is more suitable for satisfying the conservation laws, and solve the conservation equations.
The physical domain is on $y=0$ plane and the cell centers of the Cartoon meshes and cylindrical meshes overlap each other.
In addition, all geometrical, hydrodynamical, and radiation variables are defined at cell centers.

We follow the collapse and post-bounce evolution of non-rotating progenitor stars with $9.6\,{\rm M}_\odot$, $11.2\,{\rm M}_\odot$, $50\,{\rm M}_\odot$, and $70\,{\rm M}_\odot$.
The progenitor with $9.6\,{\rm M}_\odot$ (hereafter $z9.6$) is a star with zero-metallicity and is considered to be in the lowest mass range of progenitor stars of CCSNe.
Its structural feature is a steep density gradient above the iron core that enables a successful explosion even in 1D models \citep{Radice17}.
Another low mass progenitor model with $11.2\,{\rm M}_\odot$ ($s11.2$) is a widely used progenitor model with solar metalicity and taken from \cite{WHW02}.
The progenitor model with $50\,{\rm M}_\odot$ ($s50$) of \citet{Umeda08} is a blue-supergiant star with solar-metallicity and was used in the previous study of \citet{Fischer18} to explore a possible scenario of superluminous SN triggered by the QCD phase transition.
The heaviest progenitor star among our set of progenitors is $70\,{\rm M}_\odot$ ($z70$) with zero metallicity \citep{Takahashi14}.
This progenitor star may form a BH, as was explored by \citet{KurodaT18} and \citet{Shibagaki21} and is a suitable model to investigate if the QCD phase transition can turn the failed SN model into a successful one. 


In the present study we implement the EOS of \citet{Fischer18}. It features a phase transition from the DD2 nuclear relativistic mean field model of \citet{Typel10}, with density-dependent meson-nucleon couplings, to the string-flip quark matter EOS. The latter is based on a novel relativistic density functional (RDF) approach, which is build on a mechanism for (de)confinement through a string-like potential which is divergent at vanishing density. Furthermore, linear and higher-order repulsive vector interactions are included in the quark matter phase following \citet{Benic:2015}, which give rise to additional pressure contributions at high density. The RDF model has been developed originally for cold, neutron star studies by \citet{Kaltenborn17}, which was extended by \citet{Fischer18} and \citet{Bastian:2021} to finite temperatures and arbitrary isospin asymmetry, in order to be applicable to astrophysical studies of core-collapse supernovae in \citet{Fischer18} and \citet{Fischer:2020} as well as binary neutron star mergers in \citet{Bauswein19}. Finite temperature and isospin asymmetry aspects are particularly important since the conditions for the onset of the phase transition have a strong temperature and charge fraction dependence \citep[c.f. Fig.~1 and the phase diagram in Fig.~7 of][]{Fischer:2021}. Here, we employ the string-flip parametrization denoted as DD2F--RDF~1.2 \citep[see Table~I in][]{Bastian:2021}. It features densities for the onset of the phase transition of $3.25\times\rho_{\rm sat}$\footnote{The saturation density has a value of $\rho_{\rm sat}=2.5\times 10^{14}$~g\,cm$^{-3}$ or equivalent in nuclear units $0.15$\,fm$^{-3}$.} at zero temperature ($T=0$) and $1.15\times\rho_{\rm sat}$ at $T\simeq 50$\,MeV \citep[see Table~1 and Fig.~1 in][]{Fischer:2021}. The maximum mass of DD2F--RDF~1.2 for cold and non-rotating hybrid stars of 2.16\,M$_\odot$ is in agreement with the most precisely known massive pulsar of $2.08\pm0.07$\,M$_\odot$ for the PSR~J0740+6620 pulsar obtained by the analysis of \citet{Antoniadis13}, which has been recently revisited by \citet{Fonseca:2021}.
In addition, this hybrid EOS is compatible with the recent neutron star radii constraints. These are obtained by the NICER collaboration analysis for intermediate-mass neutron stars of the pulsar PSR~J0030+0451 in \citet{NICER_Miller2019} with a mass of $1.44$~M$_\odot$ and a radius of $13.02_{-1.06}^{+1.24}$\,km \citep[see also][]{NICER_Watts2019}, as well as for the pulsar PSR~J0740+6620, which is a neutron star of about 2\,M$_\odot$, by \citet{NICER_Riley2021} and \citet{NICER_Miller2021} where a radius constraint was deduced of $11.4-16.3$\,km at the 68\% confidence level.

We note further that the DD2F nuclear relativistic mean field EOS has been modified from its original DD2 version in order to be consistent with the supersaturation density constraint deduced from the heavy-ion collision data analysis of the elliptic flow by \citet{Danielewicz:2002}. For the phase transition construction from DD2F to RDF, a Maxwell-like construction is applied~\citep[further details can be found in][]{Bastian:2021}.

The 2D axially symmetric computational domain extends to $1.5\times10^4$\,km, for $z70$ and $s50$, and to $7.5\times10^3$\,km, for $s11.2$ and $z9.6$, from the center. Reason of the smaller computational domain for $s11.2$ and $z9.6$ is simply because that our EOS tabulation does not cover the required low density region. In the computational domain, nested boxes with 11(10) refinement levels are embedded for $z70$ and $s50$ ($s11.2$ and $z9.6$), and each nested box contains $64\times 64$ cells so that the finest resolution at the center becomes $\sim$115 m (a detailed resolution study can be found in Appendix~\ref{app:Spatial resolution dependence}). The neutrino energy space $\varepsilon$ logarithmically covers from 1 to 300\,MeV with 20 energy bins.

In this study, we use the up-to-date neutrino rates of \citet{Kotake18}. The electron capture rate on heavy nuclei is the most elaborate one following \citet{juoda}. Furthermore, we also take into account the following corrections: inelastic contributions and weak magnetism corrections of \citet{horowitz02} for the charged-current neutrino emission and absorption processes involving the unbound nucleons as well as for the neutral-current neutrino nucleon scattering processes; a correction for the effective nucleon mass \citep{Reddy99};  
the quenching of the axial-vector coupling constant \citep[][]{Carter&Prakash02,Fischer16}; many-body and virial corrections of \citet{horowitz17} for the neutral-current neutrino nucleon scattering rates;  effective strangeness contributions reducing the axial-vector coupling constant following \citet{horowitz02}, also here for both charged-current neutrino emission and absorption processes involving the unbound nucleons as well as for the neutral-current neutrino nucleon scattering processes.
 Following \citet{Melson15b}, we use $g_A^s=-0.1$ \citep{Hobbs16} for the effective strangeness contributions in this work, where this value increases(decreases) the cross section for $\nu p$ ($\nu n$)-scattering.

In the quark matter phase, the description of weak interaction rates based on microscopic calculations is an active field of research \citep[c.f.][and references therein]{Berdermann:2016}, e.g., exploring the role of pairing which might modify the neutrino opacity significantly in comparison to the simplistic and well-known non-interacting Fermi-gas picture.
However, such computations are very demanding and hence not feasible in simulations of astrophysical applications. Instead, we employ the standard weak rates in quark matter featuring baryons. These include the charged-current absorption reactions and the inelastic neutrino-nucleon scattering processes. Therefore, we reconstruct the baryon and charge chemical potentials, $\mu_{\rm B}$ and $\mu_Q$, from the up- and down-quark chemical potentials, $\mu_u$ and $\mu_d$, as follows,
\begin{equation}
\mu_{\rm B} = 2\mu_d + \mu_u~,\quad \mu_Q = \mu_u - \mu_d~,
\end{equation}
which in turn give rise to the neutron and proton chemical potentials, $\mu_n=\mu_{\rm B}$ and $\mu_p = \mu_{\rm B} + \mu_Q$, respectively, as well as the corresponding number densities for neutrons and protons, $n_n$ and $n_p$, are being reconstructed from the up- and down-quark number densities, $n_u$ and $n_d$,
\begin{equation}
n_n= \frac{1}{3}\left(2n_d + n_u\right)~,\quad n_p= \frac{1}{3}\left(2n_u + n_d\right)~.
\end{equation}
This simplification is justified since at the conditions where quark matter appears, neutrinos of all flavors are entirely trapped. The diffusion timescale is on the order of seconds and cannot change significantly, even if the opacity for the corresponding weak interactions with quarks should strongly decrease, since neutrinos would still be trapped through a variety of other, purely leptonic weak processes such as neutrino scattering on electrons and positrons as well as neutrino pair reactions. Only on a supernova simulation timescale on the order of seconds, when neutrinos would start to diffuse from the quark matter phase, the aforementioned approximations can no longer be applied.

\section{Results}
\label{sec:Results}
In this section, we present results of our simulations.
We begin with a description of overall dynamics of all models.
After that, we focus on a particular model $s50$ and explain its second collapse and bounce in detail.

\subsection{Collapse of stellar and PNS core}
\label{sec:Collapse of stellar and PNS core}
\begin{figure}[t!]
\begin{center}
\includegraphics[angle=-90.,width=\columnwidth]{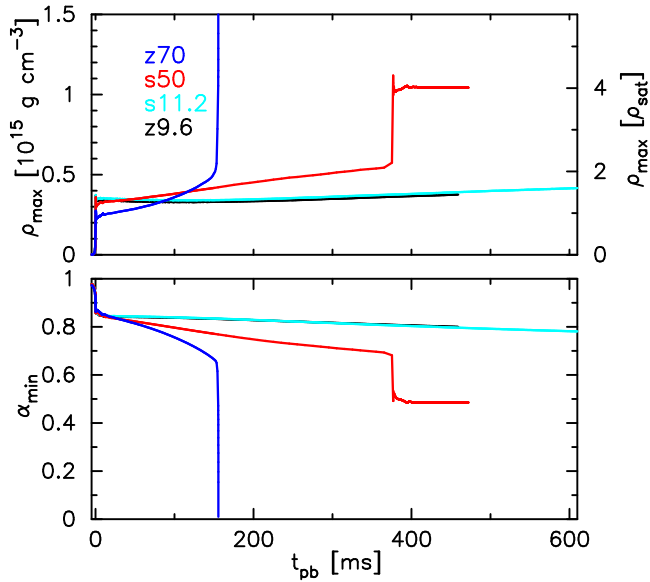}
\caption{{\em Top panel}: evolution of the maximum rest mass density for all models in units of $10^{15}$~g\,cm$^{-3}$ (left axis) and of $\rho_{\rm sat}$ (right axis). {\em Bottom panel}: evolution of the central lapse function. 
\label{fig:DmaxAmin}}
\end{center}
\end{figure}

Figure~\ref{fig:DmaxAmin} shows the evolution of maximum density and minimum lapse function for all models, from which we can gain a first glance of overall dynamics.
After the central density exceeds nuclear saturation density, the stellar core collapse halts and the core experiences the first core bounce due to stiffening of the nuclear EOS.
We define the moment of core bounce, $t_{\rm pb}=0$ when the maximum rest mass density is reached.
It corresponds to the following values for the maximum, central density of $\rho_{\rm max}=3.81$, 3.92, 3.62 and $2.73\times10^{14}$\,g\,cm$^{-3}$ for models $z9.6$, $s11.2$, $s50$ and $z70$, respectively. Roughly speaking, therefore, the maximum density at bounce has an inverse relation with the progenitor mass.
This is related to the temperatures obtained in these simulations, which are highest(lowest) for the more(less) massive progenitors, i.e. the thermal pressure additionally supports the halt of the stellar core collapse and the subsequent core bounce.

\begin{figure}[t!]
\begin{center}
\includegraphics[width=\columnwidth,angle=-90.]{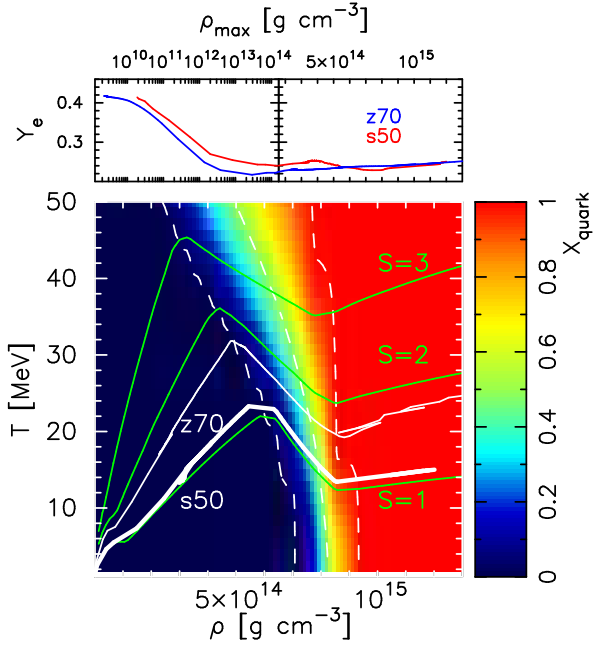}
  \caption{{\em Top panel}:~evolution of the central $Y_e$ as a function of the central rest mass density, $\rho_{\rm max}$, for the model $s50$ (red line) and $z70$ (blue line). {\em Bottom panel}:~phase diagram of the DD2F--RDF~1.2 hybrid-EOS, showing the quark volume fraction $X_{\rm quark}$ defined in \citet{Bastian:2021}, in the density-temperature plane with a fixed electron fraction of $Y_e=0.25$, which is a typical central value after bounce for both $s50$ and $z70$. The white thick(thin) line denotes the evolutionary path of quantities for $s50(z70)$. The green and white dashed lines are showing the curves of constant entropy per particle of $s=1$, 2, 3 $k_{\rm B}$ and contours of $X_{\rm quark}=0.01$ (left), 0.5 (middle) and 0.99 (right), respectively.
 \label{fig:PhaseDiagram}
}
\end{center}
\end{figure}

\begin{figure}[t!]
\begin{center}
\includegraphics[width=0.65\columnwidth,angle=-90.]{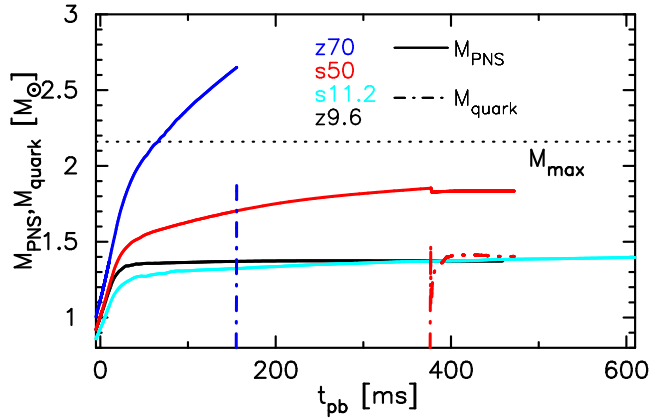}
\caption{Post-bounce evolution of the rest mass of the PNS mass $M_{\rm PNS}$ (solid lines) and proto-quark star mass $M_{\rm quark}$ (dashed) for all models. The horizontal dotted line marks the maximum mass at $T=0$ of $M_{\rm max}=2.16$\,M$_\odot$ for the current EOS. $M_{\rm quark}$ for $z9.6$ and $s11.2$ is zero since the hadron-quark phase transition region is not reached during our simulation time of $t_{\rm pb}\lesssim600$\,ms.
\label{fig:MnsMquark}
}
\end{center}
\end{figure}

After the first bounce, the maximum density again increases with time in all models due to the continuous mass accretion prior to the possible onset of a shock revival. While the less massive progenitor stars, $z9.6$ and $s11.2$, show only a slight increase compared to the other two more massive stars, the model $z70$ shows the most rapid increase rate, which features the highest mass accretion rate. Consequently, $z70$ reaches the highest central temperatures and luminosities during the post-bounce evolution (see the thin white line in Fig.~\ref{fig:PhaseDiagram} and the top panel of Fig.~\ref{fig:LnuEnu}).  At the same time, $z70$ exhibits an accelerated decrease in the minimum lapse function $\alpha_{\rm min}$, as illustrated in the bottom panel of Fig.~\ref{fig:DmaxAmin}, which decreases below $\lesssim0.01$ at $t_{\rm pb}\sim155$\,ms and we observe the formation of the apparent horizon. At our final simulation time for $z70$, i.e. less than a millisecond after the apparent horizon appears, its circumferential radius is $\sim$2.29\,km, with its corresponding mass being $\sim$2.40\,M$_\odot$. Since the initial condition is spherically symmetric, the apparent horizon essentially preserves spherically symmetric property with its deviation from spherical symmetry being less than $0.1\%$.
We note that the size of the quark core at this moment has a radius of $\sim1.8$\,km, which is well within the apparent horizon and is thus causally disconnected from outside. 
%
%
A comparison with the study of \cite{Shibagaki21}, where the same $z70$ progenitor model was used, shows the black hole formation at a significantly later post bounce time of $t_{\rm pb}\simeq 230$\,ms. This is attributed to the different, purely hadronic EOS of \citep{LSEOS} used in the supernova simulations of \cite{Shibagaki21}, which are based on 3D general relativistic neutrino-radiation hydrodynamics simulation. In their case the PNS collapse and black hole formation is induced due to enclosed PNS mass reaching the maximum mass of the hadronic EOS. In contrast here, the PNS collapse is due to reaching the critical conditions for the onset of the phase transition. The phase boundaries, as well as their temperature dependence, are marked as green dashed lines in Fig.~\ref{fig:PhaseDiagram}, for the onset of the hadron-quark mixed phase (left) and for the pure quark matter phase (right). It becomes evident that the evolution proceeds through the hadronic-quark mixed phase into the pure quark matter phase. However, when the second collapse halts in the pure quark matter phase, and even though a second shock wave has formed in compliance with the previous study of \citet{Fischer18}, the enclosed mass of about 2.7\,M$_\odot$ (see Fig.~\ref{fig:MnsMquark}) exceeds also here the maximum mass given by the DD2F--RDF~1.2 hybrid EOS and a black hole forms eventually.

The $s50$ model takes an intermediate evolutionary path between the less massive ($z9.6$ and $s11.2$) and the most massive ($z70$) progenitor stars. The maximum density increases significantly faster than the lower mass models, due to a higher post-bounce mass accretion rate, though slower than for $z70$. At $t_{\rm pb}\sim376$\,ms, when the maximum density reaches slightly above two times of the saturation density, the PNS core enters the phase transition region, as illustrated via the thick white line in Fig.~\ref{fig:PhaseDiagram}. 
Similar as for the $z70$ simulation, also for $s50$ the central PNS becomes gravitationally unstable and collapses once a sufficiently large amount of the PNS interior enters the hadron-quark mixed phase, where the EOS softens substantially in comparison to the pure hadronic and quark matter phases at lower and higher density, respectively.
During this second collapse, the maximum density increases rapidly from $\rho_{\rm max}\simeq 2\times \rho_{\rm sat}$ to slightly above $\rho_{\rm max}\gtrsim 4\times\rho_{\rm sat}$ (see Fig.~\ref{fig:DmaxAmin}).
At such a high density, baryons are completely dissociated into quarks and the stiffness of EOS again increases mediated by the repulsive vector interactions.
Consequently the NS collapse halts and the second core bounce occurs at $t_{\rm pb}=376.6$\,ms. The second bounce time seen in the model $s50$ is considerably earlier than the finding of about $1.2$~s in \citet{Fischer18}. 
We attribute the reason therefore to the different hybrid EOS applied in \citet{Fischer18}, corresponding to the DD2--RDF~1.1 nomenclature of \citet{Bastian:2021}. It features a generally higher onset density for the hadron-quark phase transition, on the order of $3.5\times\rho_{\rm sat}$ at $T=0$, as well as a different temperature dependence. Further differences in the simulations may arise due to the different hydrodynamics scheme of of the supernova model of \citet{Fischer18}, which is based on an adaptive Lagrangian mass grid and differs from the one in this study. Furthermore, we have checked if the spatial resolution can influence on the evolution up to and including second bounce and found a rough numerical convergence, which is discussed in Appendix~\ref{app:Spatial resolution dependence}.

Figure~\ref{fig:PhaseDiagram} depicts how the central PNS evolves in the phase diagram of the DD2F--REF~1.2 hybrid EOS. The top panel shows the evolution of the central $Y_e$ as a function of the central rest mass density for the models $s50$ (red line) and $z70$(blue), with $Y_e\simeq 0.25$ for both simulations. The bottom panel exhibits the phase diagram of hybrid EOS employed with the color-coding indicating the quark volume fraction, $X_{\rm quark}$, in the density-temperature plane for a fixed $Y_e=0.25$. The white thick and thin solid lines denote evolutionary paths of the central density and temperature for $s50$ and $z70$, respectively. The green lines show curves of constant entropy per particle of $s=1$, 2, 3 $k_{\rm B}$, and the vertical white dashed lines correspond to $X_{\rm quark}=0.01$, 0.5 and 0.99.

In both models $s50$ and $z70$, the central region evolves nearly along a constant entropy path after the first bounce due to the adiabatic compression. It lasts until it reaches the purely quark matter phase, i.e., where $X_{\rm quark}=1$. From the phase diagram, we clearly see a temperature dependence of the initiation of the hadron quark mixed phase, i.e. the left vertical white dashed curve. It shows that the mixed phase appears at lower density for matter with higher temperature or entropy. This turns out that the model $z70$, possessing higher core entropy, enters the phase transition region at a slightly lower density of $\rho_{\rm max}\simeq5\times10^{14}$\,g\,cm$^{-3}$, compared to the value of $\rho_{\rm max}\simeq6\times10^{14}$\,g\,cm$^{-3}$ in $s50$.
After the central region enters the pure quark matter phase, we witness multiple core bounces for $z70$, due to the stiffening of EOS that produce a standing shock front.
However, the mass accretion rate is extremely high, on the order of $(\sim$4\,M$_\odot$~s$^{-1}$, measured on the shock surface at the second collapse, and the shock front cannot propagate outward to blow off the outer envelop. Instead $\rho_{\rm max}$ increases continuously and a black hole is quickly formed as already explained.
On the other hand, the maximum density for $s50$ does not continuously increase once its PNS interior transforms into a quark matter core. In Fig.~\ref{fig:DmaxAmin}, $\rho_{\rm max}$ shows a single bounce and then quickly settles into a new value of $\rho_{\rm max}\simeq1.05\times10^{15}$\,g\,cm$^{-3}$ after a rapid ring down phase. At the same time, the minimum lapse $\alpha_{\rm min}$ also decreases from $\sim0.7$ to $\sim0.5$ across the second core bounce. The second bounce creates an energetic hydrodynamic shock wave due to the stiffening of EOS. In contrast to $z70$, which has a significantly higher mass accretion rate, the shock front propagates outward with a relativistic velocity of about two-thirds of the speed of light,  because of a lower mass accretion rate of about 0.33\,M$_\odot$~s$^{-1}$ measured on the shock surface at the second core collapse. The shock evolution will be discussed in the next section \ref{sec:Explosion dynamics}. 

Qualitative difference between the BH formation model ($z70$) and the explosion model ($s50$) can be explained from evolution of PNS mass. In Fig.~\ref{fig:MnsMquark}, we plot the PNS mass $M_{\rm PNS}$ (solid lines) and proto-quark star mass $M_{\rm quark}$ (dashed) for all models.
The latter is defined via the quark volume fraction exceeding one percent. It has been introduced in \citet{Fischer18} and further details, including the phase transition construction, can be found in \citet{Bastian:2021}.
We define the PNS surface by the region where the rest mass density drops below $10^{10}$~g\,cm$^{-3}$. 

In contrast to the $z70$ model, the baryon mass enclosed inside the PNS for $s50$ of about $1.83$\,M$_\odot$ is well below the maximum gravitational mass of DD2F--RDF~1.2 after the hadron-quark phase transition (see Fig.~\ref{fig:MnsMquark}). Note that the maximum allowed mass for hot neutron stars can be increased by $\sim$0.2--0.4\,M$_\odot$, compared to their corresponding cold case due to the corresponding additional pressure support from hot matter \citep[c.f.][]{Hempel12}. The nascent PNS for $s50$ features a massive quark matter core of about 1.43\,M$_\odot$. Similar was the case for $z70$, a strong second hydrodynamics shock wave forms as a result of the phase transition. However, in the case of $s50$ the remnant PNS is semistable, e.g., with respect to radial and non-radial perturbations such that a black hole formation is not observed. The dynamics of the second shock wave has important consequences for the supernova explosion dynamics, which will be discussed below.

\begin{table*}
\centering
\begin{tabular}{ccccccc} 
\hline
  model 
  & $\xi_{1.5}^{\rm b(i)}$
  & $\xi_{2.5}^{\rm b(i)}$
  & $M_{\rm remnant}$
  & $M_{\rm quark}$ 
  & $t_{\rm exp}$
  & $t_{\rm final}$ 
  \\ 
  & 
  & 
  & $[{\rm M}_\odot]$
  & $[{\rm M}_\odot]$
  & $[$ms$]$ 
  & $[$ms$]$ 
  \\ 
\hline
\hline
$z9.6$ & $(2.2\times10^{-4})$ & $(7.6\times10^{-5})$& 1.37 & --- & 167 & 459  \\ 
\hline
 $s11.2$ & $2.4(1.8)\times10^{-1}$ & $(5.0\times10^{-3})$ & 1.40 & --- & 503 & 650  \\ 
\hline
$s50$ & 2.3(1.6)& 1.9(1.9)$\times10^{-1}$& 1.83 & 1.35 & 384 & 450  \\ 
 \hline
$z70$ & 4.0(1.1) & 1.6(1.0) & 2.65 & 1.88 & ---  & 155  \\ 
 \hline
\end{tabular}\\
\caption{List of models and their simulation results, showing the compactness parameters, $\xi_M$, measured at at two different mass coordinates of 1.5\,M$_\odot$ and 2.5\,M$_\odot$, the remnant masses, $M_{\rm remnant}$, the mass of the quark-matter core, $M_{\rm quark}$, the explosion time, $t_{\rm exp}$, and the final simulation time, $t_{\rm final}$.
We evaluate $\xi_M$ at core bounce and at the initial data, denoted by $\xi_M^{\rm b}$ and $\xi_M^{\rm i}$ (values in parantheses), respectively.
Note that our calculation domain does not contain $1.5\,{\rm M}_\odot$ for $z9.6$ and $2.5\,{\rm M}_\odot$ for $s11.2$ and we omit the corresponding compactness parameters at bounce $\xi_M^b$ from the table.
Explosion time $t_{\rm exp}$ denotes the post bounce time when the average shock radius reaches $r=1000$\,km.
\label{tab:model_list}}
\end{table*}

None of the lighter progenitor models, $z9.6$ and $s11.2$, reach the conditions required for the hadron-quark phase transition in our simulation times of less than $\sim$600\,ms after bounce. Instead, these two models show the explosion driven by neutrino heating.
This will be explained in details in the next sec.~\ref{sec:Explosion dynamics}. It is noteworthy that the maximum density continues increasing slightly after the explosion begins. If the central density increases with the same rate at the final simulation time $\dot\rho_{\rm max}\sim10^{14}$~g\,cm$^{-3}$~s$^{-1}$, we estimate that it will exceed $2\times\rho_{\rm sat}$ at $t_{\rm pb}\sim1.5$ s in both models and the hadron-quark phase transition may occur during the subsequent deleptonization phase of the nascent NS. However, we note here also that the phase transition onset mass of DD2F--RDF~1.2 for cold neutron stars, i.e. the mass threshold for the hybrid star branch lies above 1.37\,M$_\odot$. Hence, in case the phase transition would occur we expect that only a tiny fraction, if any, of the neutron star's interior would consist of pure quark matter.

In Table~\ref{tab:model_list} we list a summary of progenitor models and their final fates.
The compactness parameter $\xi_M$ measures the ratio of mass $M$(M$_\odot)$ and radius $R_M$, where the latter term $R_M$ expresses the radius, which encloses the mass $M$ \citep{O'Connor11}, and is in units of 1000 km.
Regarding the compactness of progenitor star, $\xi$, $z70$ shows the highest value. \cite{O'Connor11} discussed a relation between the bounce compactness $\xi^{\rm b}_{2.5}$ and BH formation time. According to their result, BH formation time is less than a few 100\,ms for very massive stars having $\xi^{\rm b}_{2.5}\gtrsim1.5$, which is in accordance with the current model $z70$. $s50$ forms a hybrid star with its quark core and total PNS mass reach $1.43$ and $1.83$\,M$_\odot$, respectively. $z9.6$ and $s11.2$, which have a small compactness parameter, show almost the same PNS mass $1.37$\,M$_\odot$ with their evolution showing a plateau behavior (see black and cyan lines in Fig.~\ref{fig:MnsMquark}).

\begin{figure}[t!]
\begin{center}
\includegraphics[width=0.975\columnwidth,angle=-90.]{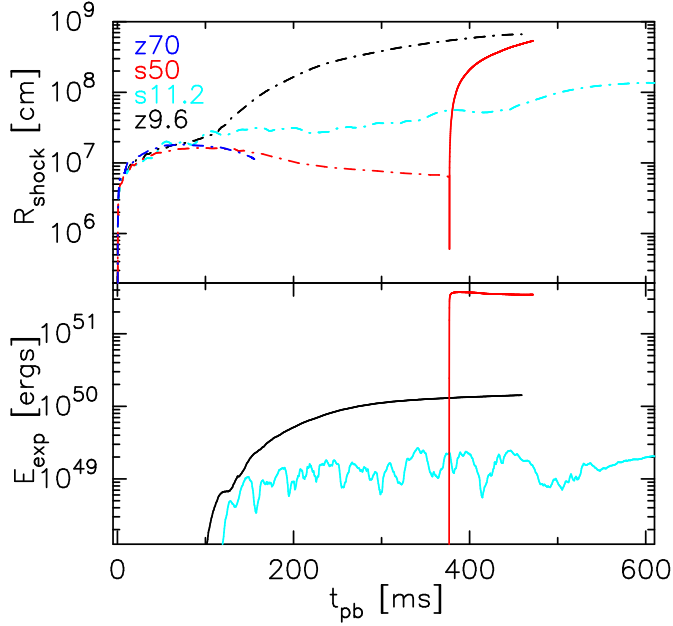}
\caption{Post-bounce evolution of the averaged shock radii $R_{\rm shock}$ (top panel) and of the diagnostic explosion energy $E_{\rm expl}$ (bottom panel) for all models. In the upper panel, we distinguish the core bounce shock (denoted by the dash-dotted line) and the second bounce shock (solid).
\label{fig:RshockEexp}
}
\end{center}
\end{figure}

\subsection{Explosion dynamics}
\label{sec:Explosion dynamics}
In this section we discuss the explosion dynamics for models $z9.6$, $s11.2$, and $s50$. As $z70$ does not explode, its $E_{\rm exp}$ is zero. Fig.~\ref{fig:RshockEexp} exhibits time evolution of averaged shock radii $R_{\rm shock}$ (top panel) and diagnostic explosion energy $E_{\rm exp}$ (bottom).
We note that $E_{\rm exp}$ does not take into account the binding energy of the stellar envelop. In addition nuclear recombination and the associated nuclear burning energy generation are also neglected. 

We begin by $z9.6$. From its shock evolution (black curve in Fig.~\ref{fig:RshockEexp}), we see that the shock expands slowly at initial times up to about 100\,ms post bounce.
After about 120\,ms post bounce, this model enters a shock runaway phase.
The shock front reaches 1000\,km at $t_{\rm pb}=167$\,ms. At the same time, the diagnostic explosion energy $E_{\rm exp}$ starts increasing and eventually reaches a value of $1.4\times10^{50}$\,erg when the shock front reaches the outer boundary of the calculation domain.
In \citet{Melson15a} and \citet{Radice17}, the authors also performed multi-D SN simulations using the same $9.6$\,M$_\odot$ progenitor star. Although the detailed numerical setups are different in each study, e.g., different hydrodynamics schemes, grids, neutrino transport and weak rates, the shock and diagnostic explosion energy evolution show quantitative behavior.
\cite{Melson15a} and \cite{Radice17} report $E_{\rm exp}\sim8\times10^{49}$\,erg and $\sim1.1\times10^{50}$\,erg, respectively, at $t_{\rm pb}=400$\,ms, where the latter study takes into account the gravitational binding energy of the envelope, which is merely $-2\times10^{48}$\,erg \citep[c.f.][]{Melson15a}.
Our diagnostic explosion energy is within a range of uncertainty due to different numerical setups and we thus find consistent results with previous studies for model $z9.6$.

\begin{figure*}[t!]
\begin{center}
\includegraphics[width=0.95\columnwidth,angle=-90.]{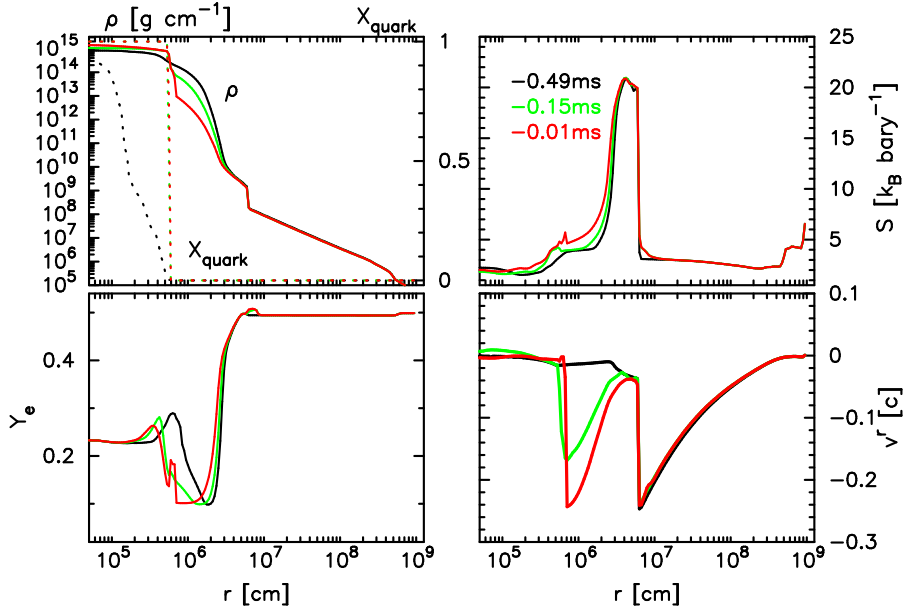}
  \caption{Radial profiles of the rest mass density $\rho$ (solid lines) and quark fraction $X_{\rm quark}$ (dotted), entropy per particle $s$, electron fraction $Y_e$, and radial component of the three-velocity $v^r\equiv u^r/u^t$ for the model $s50$ at different times before to the second bounce, denoted as $t_{\rm p2b}$, corresponding to the onset and proceeding collapse of the PNS due to the hadron-quark phase transition.
  \label{fig:QcbM50bb}
}
\end{center}
\end{figure*}

For the other low mass model $s11.2$, the average shock radius steadily increases without any visible shock stagnation, similarly to $z9.6$, however, at a somewhat slower rate. This is attributed to the extended silicon-sulphur layer for this progenitor model, which is absent for $z9.6$. It results in higher mass accretion rates post bounce for $s11.2$, which, in turn slow down the neutrino driven shock expansion.
Fig.~\ref{fig:RshockEexp} shows that the mean shock radius $R_{\rm shock}$ reaches 1000\,km at $t_{\rm pb}\sim503$\,ms and the diagnostic explosion energy reaches $3$-$4\times10^{49}$\,erg at the end of simulation. These values are in quantitative agreement with those in \citet{BMuller15b}, in which they conducted a relativistic 2D CCSN simulation for the same 11.2\,M$_\odot$ progenitor model. Although the employed EOS and neutrino opacities are different, and comparison is not straightforward, his corresponding 2D model shows a similar evolution with $s11.2$ in this study.
The neutrino driven supernova explosion observed for this low mass model $s11.2$ is in qualitative agreement with the results reported by other groups based on axially symmetric simulations \citep[c.f.][]{Buras06a,Marek09,Takiwaki14,Nagakura18}. More controversial are the recent 3D supernova simulations of this progenitor model \citep[e.g.,][]{Takiwaki12,Tamborra14,takiwaki16}, which is an active subject of research.

\begin{figure*}[t!]
\begin{center}
\includegraphics[width=0.95\columnwidth,angle=-90.]{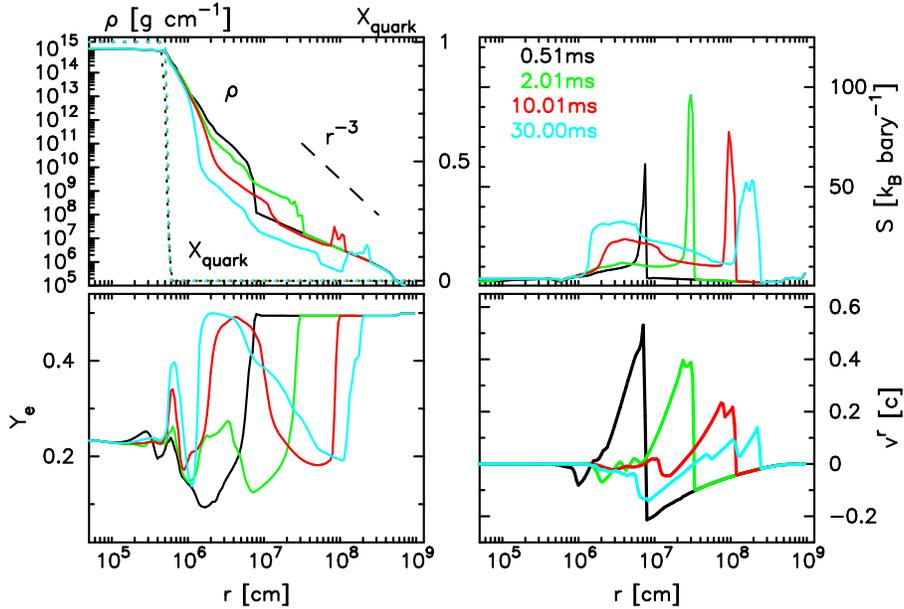}
  \caption{The same as Fig.~\ref{fig:QcbM50bb}, but for different time slices after the second bounce. In the density profile, we show a slope of $r^{-3}$ by a dashed line.
\label{fig:QcbM50ab}}
\end{center}
\end{figure*}

In comparison to these less massive progenitor stars, $s50$ shows a continuous shock recession once the shock stagnates around $t_{\rm pb}\sim100$\,ms. The recession lasts until the second collapse occurs (see Fig.~\ref{fig:RshockEexp}). The red line in the top panel shows that the averaged shock radius eventually shrinks to $R_{\rm shock}=62$\,km. Then concomitantly with the second bounce, a strong, second shock wave is produced at the surface of quark core, locating at around $r\sim7$\,km. This shock front overtakes the preceding standing shock front at $r=62$\,km and propagates further out with a relativistic velocities, i.e. within a few ms after the second bounce, the shock front exceeds $r=1000$\,km. Simultaneously the diagnostic explosion energy rises continuously, and reaches its asymptotic value of $E_{\rm exp}\sim3.5\times10^{51}$\,erg.
Afterward, $E_{\rm exp}$ shows nearly a constant value throughout the runaway explosion phase.
We estimate that the gravitational binding energy of the envelop is roughly $-4.0\times10^{50}$\,erg and thus the asymptotic explosion energy would be around $3\times10^{51}$\,erg.
In view of the explosion dynamics, similar as was discussed in \citet{Fischer18}, our model $s50$ supports that these massive progenitor stars could be the progenitors of a kinds of superluminous supernova events, , \citep[c.f. the discussion in][and references therein]{Smith07,Moriya13}, if a sufficiently strong hadron-quark phase transition occurs.


Next, to illustrate the dynamics at the second collapse and in the subsequent explosion phase, we plot spherically averaged profiles of the rest mass density $\rho$ and quark fraction $X_{\rm quark}$ (top left panel), entropy $s$ (top-right panel), electron fraction $Y_e$ (bottom left panel), and the radial component of the three-velocity $v^r$ (bottom right panel) in Figures~\ref{fig:QcbM50bb} and \ref{fig:QcbM50ab}, at selected time slices.

Fig.~\ref{fig:QcbM50bb} shows profiles before second bounce $t_{\rm p2b}<0$, where $t_{\rm p2b}$ denotes the post second bounce time and written in the top right panel.
From the density and radial velocity profiles, we see that the second collapse takes place within a fraction of a millisecond.
At $t_{\rm p2b}=-0.49$\,ms (black lines), the radial velocity profile shows nearly a hydrostatic profile ($v_r\sim0$) with a slight infalling sign inside the PNS.
At the same time, quarks are already partially deconfined, showing $X_{\rm quark}\sim0.9$ at the center.
We note that quarks are completely confined in hadronic-matter, i.e. $X_{\rm quark}=0$, up until $\sim1$\,ms before the second bounce, which is a major difference from \cite{Zha20}.
Shortly after that, the PNS core collapses with a relativistic speed of $\sim0.25c$ (see the red line in $v^r$ profile) and the PNS core transforms into a quark core with $X_{\rm quark}=1$.
While the entropy (top-right) and $Y_e$ (bottom-left) do not show any noticeable changes compared to $X_{\rm quark}$ and $v^r$.

After the second bounce happens, the hydrodynamic profiles change considerably.
The radial velocity profile shows a relativistic outward shock propagation immediately at $t_{\rm p2b}=0.51$\,ms and,
already at this moment, the entropy behind the shock reaches $s\ge60$ $k_{B}$ baryon$^{-1}$.
In the subsequent phase, however, its propagation speed decreases gradually as clearly be seen in the bottom right panel.
At $t_{\rm p2b}=30$\,ms, the shock decelerates to $\sim0.1c$ (cyan line in the bottom-right panel).
The entropy behind the shock also decreases noticeably from $\sim100$ $k_{B}$ baryon$^{-1}$ at $t_{\rm p2b}=2$\,ms to $\sim50$ $k_{B}$ baryon$^{-1}$ at $t_{\rm p2b}=30$\,ms.
Furthermore a density bump is formed around $r\sim10^8$\,cm at $t_{\rm p2b}=10$\,ms.
All of these facts indicate that the shock front propagates in a dense low-entropy medium and is substantially decelerated.
Indeed from the top left panel in Fig.~\ref{fig:QcbM50ab}, the angle averaged density profile shows a shallower density gradient than the slope of $r^{-3}$ at $10^7$\,cm $\lesssim r \lesssim$ $5\times10^8$\,cm.
In such a dense environment, the shock can be decelerated very efficiently \citep{Sedov1959,Wongwathanarat15}, resulting in development of Rayleigh-Taylor (RT) instabilities, which we will describe later.

The $Y_e$ profile also shows a remarkable feature.
The low-$Y_e$ materials, which initially locate in the $Y_e$-trough just above the PNS surface $r\sim20$\,km (see black line in the bottom-left panel), are ejected along with the explosion, since the explosion initiates on the quark core surface at $r\sim7$\,km, i.e. underneath the $Y_e$-trough.
At $t_{\rm p2b}=30$\,ms, those low-$Y_e$ ejecta are reaching $r\sim10^8$\,cm and produce a steep $Y_e$ composition gradient across the shock surface.
We also find a high-$Y_e$ and high entropy region, for instance, with $s\sim30$ $k_B$ baryon$^{-1}$ and $Y_e\sim0.5$ around $r\sim5\times10^6$\,cm at $t_{\rm p2b}\ge30$\,ms.
These components are produced due to a powerful neutrino irradiation.

\begin{figure*}[t!]
\begin{center}
\includegraphics[width=0.95\columnwidth,angle=-90.]{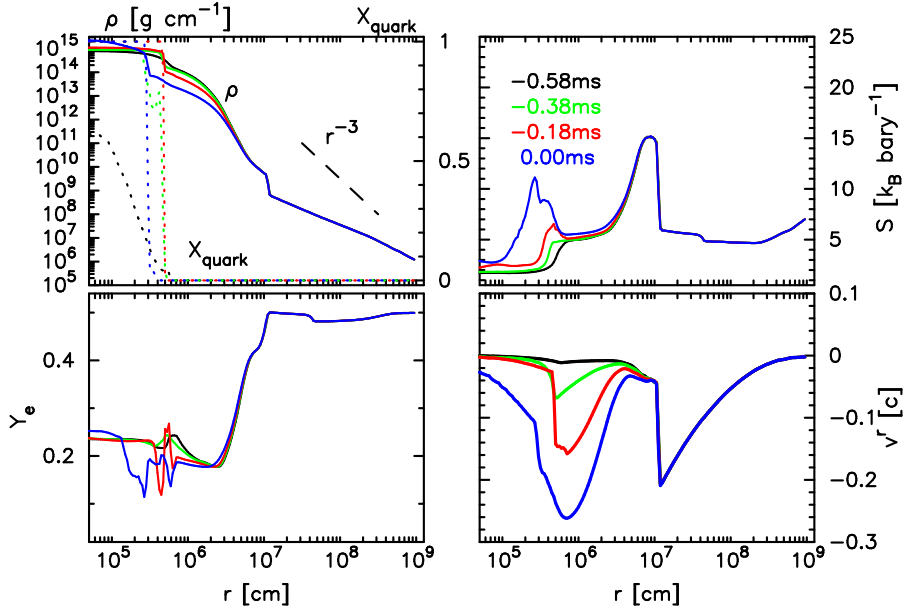}
  \caption{The same as Fig.~\ref{fig:QcbM50ab}, but for model $z70$ at several time slices relative to the final simulation time denoted in the top-right panel.
  The circumferential radius of AH is $r\sim2.29$\,km at the final simulation time.
\label{fig:QcbM70}}
\end{center}
\end{figure*}
It is informative to show the second ``bounce'' in model $z70$, as it is associated with the BH formation.
Fig.~\ref{fig:QcbM70} is the same as Fig.~\ref{fig:QcbM50ab}, but for model $z70$.
Time is measured relative to the final simulation time, i.e. $t-t_{\rm final}$, and denoted in the top-right panel.
At $t-t_{\rm final}=-0.58$\,ms, the highest value for $X_{\rm quark}$ is $\sim0.6$ and thus the quark core is still not formed.
The inner core is showing the initiation of second collapse as can be seen from the black line in the bottom-right panel, which marginally shows negative radial velocities at $r\lesssim5$\,km.
Afterward the second collapse accelerates and the quark core is immediately formed.
At $t-t_{\rm final}=-0.38$ and $-0.18$\,ms (green and red lines), the shock surface is formed at $r\sim5$\,km as well as the quark core surface.
In contrast to the less massive progenitor model $s50$, the second collapse cannot be halted even after the quark core formation, which supplies an additional pressure support.
Both the shock and quark core surfaces continuously shrink with no mass ejection (blue line in the velocity profile).
At the final simulation time, the shock front recedes to $r\sim2.3$\,km, where the AH appears.
At this moment, the quark core surface locates at $r\sim3$\,km, which is above the AH (see the blue dotted line in the top left panel).

\begin{figure*}[htbp]
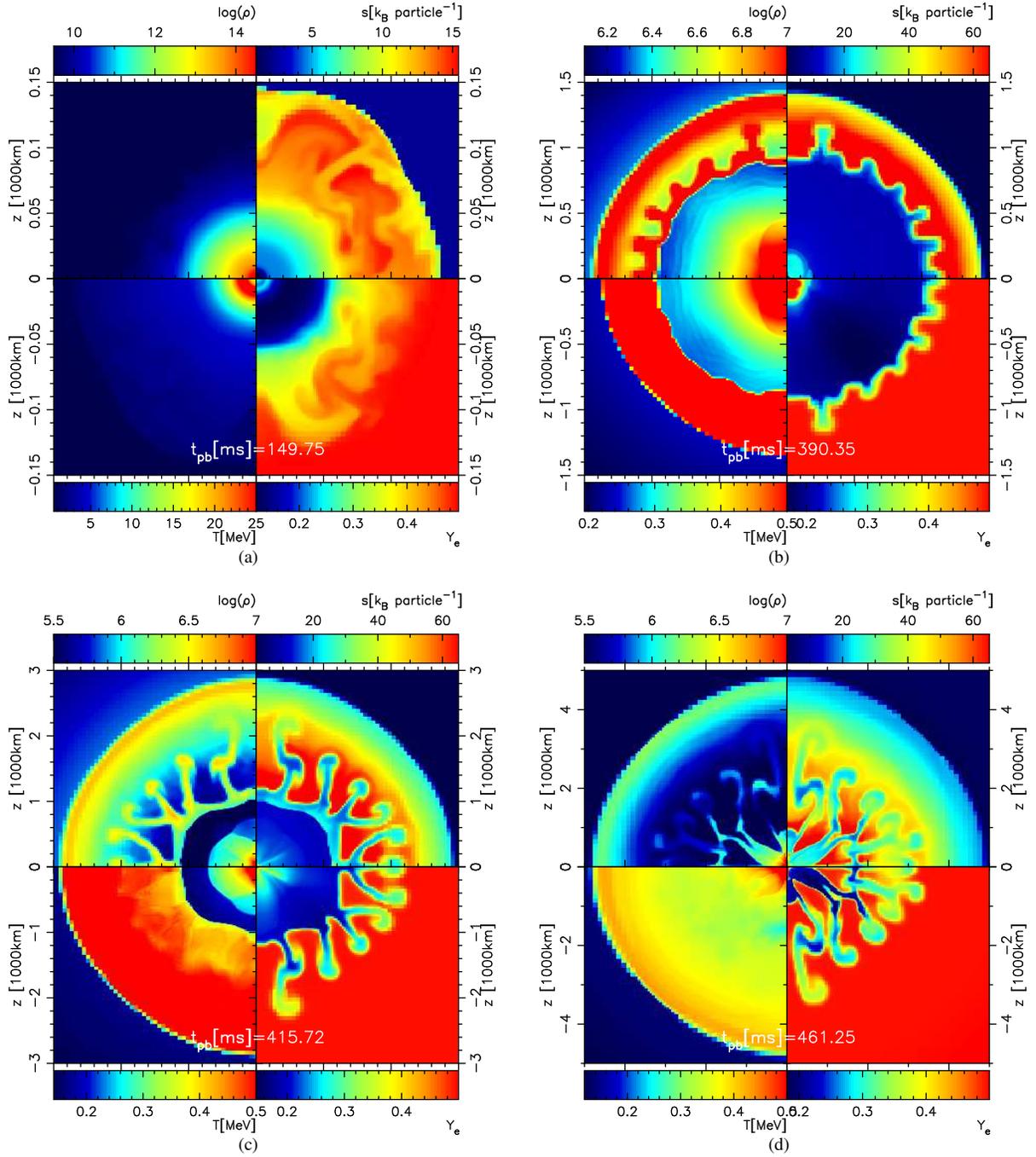

\begin{center}
\subfigure[]{\includegraphics[width=\columnwidth,angle=-90.]{cM50_1.eps}\label{fig:cM50_2D_a}}
\hspace{5mm}
\subfigure[]{\includegraphics[width=\columnwidth,angle=-90.]{cM50_2.eps}\label{fig:cM50_2D_b}}
\\
\subfigure[]{\includegraphics[width=\columnwidth,angle=-90.]{cM50_3.eps}\label{fig:cM50_2D_c}}
\hspace{5mm}
\subfigure[]{\includegraphics[width=\columnwidth,angle=-90.]{cM50_4.eps}\label{fig:cM50_2D_d}}\\
\caption{Snapshots of the $x-z$-plane for the logarithmic density log$_{10}(\rho$~[g\,cm$^{-3}$]) (top-left panels), entropy per particle $s$ [$k_{\rm B}$] (top-right panels), temperature $T$ [MeV] (bottom-left panels), and $Y_e$ (bottom-right panel) for $s50$, at different time slices $t_{\rm pb}=150$\,ms in graph~(a) as a representative of the normal structure during the post bounce phase prior to the hadron-quark phase transition as well as at $t_{\rm pb}(t_{\rm p2b})=390(14)$ in graph~(b), 416(40) {in graph~(c)} and 461(85)\,ms in graph~(d) after the second bounce. 
\label{fig:cM50_2D}
}
\end{center}
\end{figure*}
A combination of the aforementioned shock deceleration and steep gradient of $Y_e$ results in a very remarkable multi-D hydrodynamic feature.
To illustrate it, we depict 2D profiles of the shocked region at several time slices in Fig.~\ref{fig:cM50_2D}.
We select four time slices: $t_{\rm pb}\simeq 150$\,ms as a representative of normal post-bounce structure in Fig.~\ref{fig:cM50_2D_a} and $t_{\rm pb(p2b)}=390(14)$ in Fig.~\ref{fig:cM50_2D_b}, 416(40) in Fig.~\ref{fig:cM50_2D_c}, and 461(85)\,ms in Fig.~\ref{fig:cM50_2D_d} as example after the second bounce.
Each panel consists of four panels, in which we plot: logarithmic density log$_{10}(\rho$~[g\,cm$^{-3}$]) (in top-left mini panel), entropy per particle $s$ [$k_{\rm B}$] (top-right), temperature $T$~[MeV] (bottom-left), and electron fraction $Y_e$ (bottom-right) for model $s50$.

Top-left panel shows normal convective motions driven mainly by neutrino heating.
At $t_{\rm pb}=150$\,ms, the shock front locates $r\sim140$\,km and high entropy ($s\sim15$ $k_{\rm B}$ baryon$^{-1}$) convective bubbles are seen behind it.
In the post second bounce phase, the shock expands initially with a relativistic speed, though it starts decelerating already at $t_{\rm p2b}\gtrsim10$\,ms.
Indeed from the density contour at $t_{\rm pb(p2b)}=390(14)$\,ms (top-right panel), we see a dense shell locating between $10^8$\,cm $\lesssim r\lesssim1.5\times10^8$\,cm.
Within the shell, its rest mass density exceeds $\rho=10^7$\,g\,cm$^{-3}$, which is approximately one order of magnitude larger than the value of its surrounding.
Particularly interesting feature is seen on the inner edge of the shell at $r\sim10^8$\,cm, where the density has a negative radial gradient.
The deceleration of the dense shell produces an effective outward acceleration in the comoving frame, acting on the shell component.
Therefore it fulfills the condition for development of the Rayleigh-Taylor instability.
As a consequence, higher density materials, which initially locate more inward, penetrate into the lower density side and the well known mushroom like structures are formed.
Those mushroom like plumes are composed of neutron rich material with $Y_e\lesssim0.3$ (see bottom-right minipanel at $t_{\rm pb(p2b)}=390(14)$\,ms).

From Figs.~\ref{fig:cM50_2D_c} and \ref{fig:cM50_2D_d}, one can find how effectively those plumes, associated with the Rayleigh-Taylor instability, transport the low-$Y_e$ material outward.
The low-$Y_e$ plumes are growing at $r\gtrsim10^8$\,cm.
At such a distant region from the central hot PNS, neutrino irradiation is significantly weak and neutrino absorption basically does not occur.
Consequently $Y_e$ of the ejecta does not basically change except through the hydrodynamic mixing process.
The radial three-velocity of the plumes is roughly $v^r\sim0.1c$ at our final simulation time of $t_{\rm pb(p2b)}=475(100)$\,ms and approaches nearly a steady state value.
\cite{Fischer20} reported $r$-process nucleosynthesis calculation based on spherically symmetric explosion models triggered by the PDE mechanism.
According to their results, the explosion can indeed transport low-$Y_e$ material to the distant region $r\gtrsim10^{8-9}$\,cm, though the $Y_e$ value tends to become higher $\gtrsim0.35$ compared to this study.
This is simply due to the absence of the RT mixing under the spherically symmetry.
Therefore, with the aid of the RT mixing, the low-$Y_e$ components can be more efficiently ejected and may significantly influence on the nucleosynthetic yields, which will be reported elsewhere.

\begin{figure}[htbp]
\begin{center}
\includegraphics[width=55mm,angle=-90.]{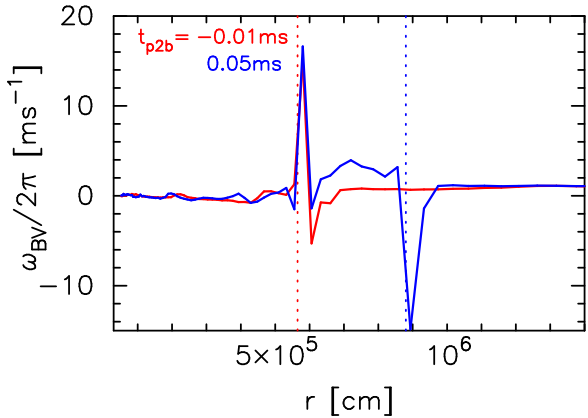}
\caption{Spherically averaged spatial profiles of the Brunt-V\"ais\"al\"a frequency $\omega_{\rm BV}$ for $s50$ around the second bounce. Here we focus only on the region above of the quark-matter core and we select two representative time slices at $t_{\rm p2b}=-0.01$\,ms (red line) and $t_{\rm p2b}=0.05$\,ms (blue line).
The vertical dashed lines indicate the averaged radii of the second bounce shock.
Negative $\omega_{\rm BV}$ indicates convectively unstable region based on the Ledoux criterion.
\label{fig:BVfrequency}}
\end{center}
\end{figure}
The Rayleigh-Taylor instability can be seeded not only by the inhomogeneous structure behind the standing shock surface (see the top-left panel at $t_{\rm pb}\sim150$\,ms in Fig.~\ref{fig:cM50_2D}), but also by the convective motions inside the quark core itself.
To explain this, we evaluate the Brunt-V\"ais\"al\"a frequency $\omega_{\rm BV}$ and plot its spherically averaged spatial profile around the second bounce in Fig.~\ref{fig:BVfrequency}.
Here, we follow the same manner as described in \cite{BMuller13} to evaluate $\omega_{\rm BV}$ (see their Eq.~(14)) in relativistic form as follows:
\begin{eqnarray}
\label{eq:BV1}
{\omega_{\rm BV}}^2=\frac{\alpha }{\rho h \phi^4}\frac{\partial \alpha}{\partial r}\left(\frac{\partial \rho(1+\epsilon)}{\partial r}-\frac{1}{c^2_s}\frac{\partial P}{\partial r}\right).
\end{eqnarray}
Using these expressions, the region with positive and negative ${\omega_{\rm BV}}^2$ corresponds to the convectively stable and unstable region, respectively.
In Fig.~\ref{fig:BVfrequency}, we plot $\omega_{\rm BV}/2\pi$ with slightly modifying the expression Eq.~(\ref{eq:BV1}) as $\omega_{\rm BV}={\rm sgn}({\omega_{\rm BV}}^2)\sqrt{|{\omega_{\rm BV}}^2|}$, so that the new $\omega_{\rm BV}$ becomes a real number with its positive and negative values indicating the convectively stable and unstable regions, respectively.
In the figure, we select two time slices across the second bounce, represented by red ($t_{\rm p2b}=-0.01$\,ms) and blue ($t_{\rm p2b}=0.05$\,ms) lines.
In these time snapshots, the quark core surface, where $X_{\rm quark}=1$, locates at $r\sim6\times10^5$\,cm, without changing its location.
While the shock front propagates from $r\sim6\times10^5$\,cm at $t_{\rm p2b}=-0.01$\,ms to $\sim9\times10^5$\,cm at $t_{\rm p2b}=0.05$\,ms.
It is obvious that $\omega_{\rm BV}$ in front of the second bounce shock shows negative, indicating convectively unstable region from the Ledoux criterion, and high angular frequencies, reaching $|\omega_{\rm BV}|/2\pi\sim10$\,ms$^{-1}$.
The convectively unstable region corresponds to the region with a steep negative lepton number gradient (see $Y_e$ profile in Fig.~\ref{fig:QcbM50bb}) as well as with the entropy jump at the second bounce shock surface.
Due to such a high $\omega_{\rm BV}$, the convective motions can fully develop behind the shock even within a millisecond and be the seed of the subsequent Rayleigh-Taylor instability.
Furthermore, in the convectively stable region (i.e. $\omega_{\rm BV}>0$) between the quark core and the second bounce shock surface ($6\times10^5$\,cm$\lesssim r\lesssim9\times10^5$\,cm at $t_{\rm p2b}=0.05$\,ms), we see the angular frequencies (or the so-called ``plume frequencies'' \citep{Murphy09}) reaching $\omega_{\rm BV}/2\pi\sim$ a few kHz.
This buoyancy frequency can reasonably explain the GW frequency at the second bounce, which will be discussed in the next section.

\begin{figure*}[t!]
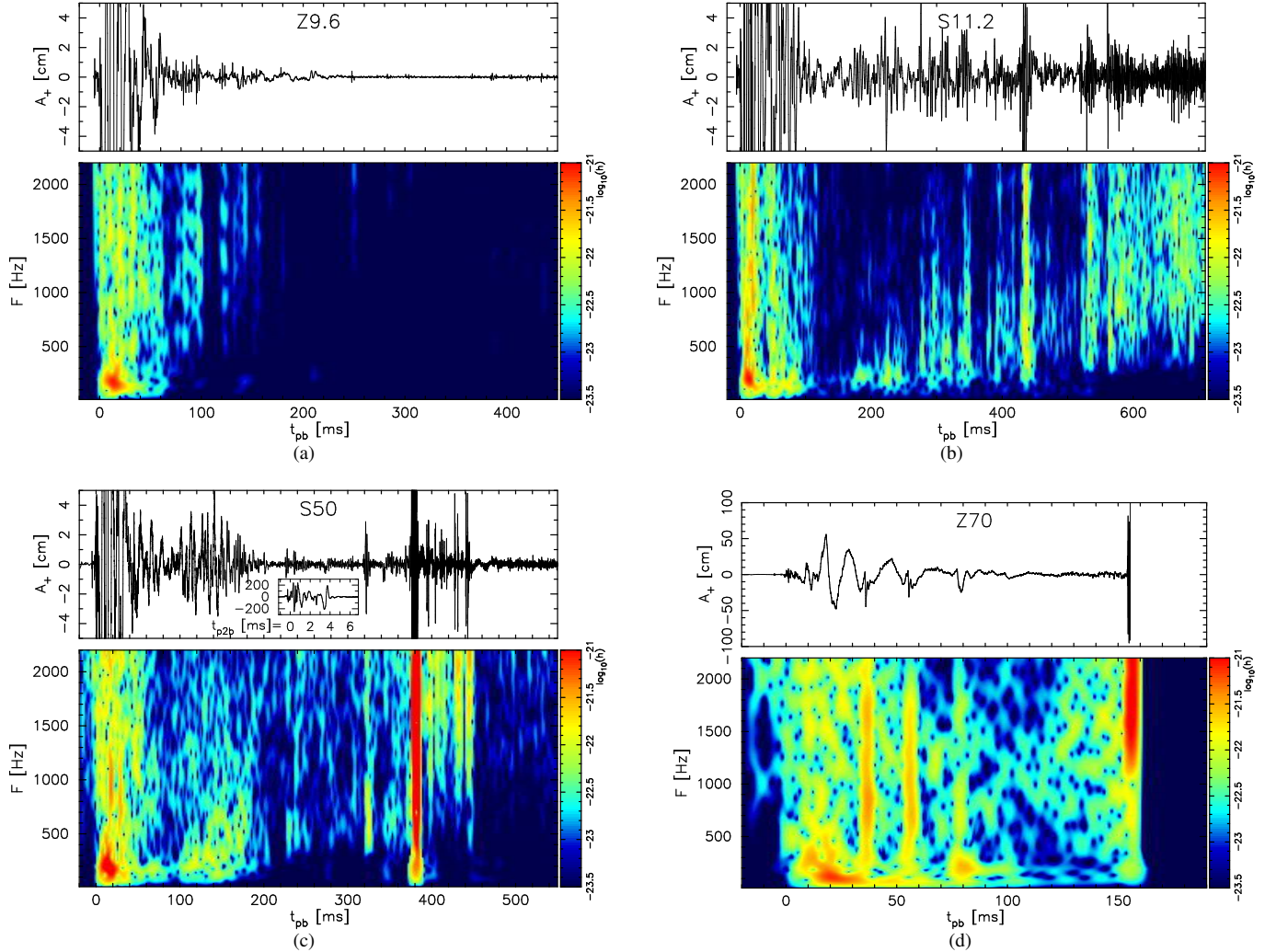

\centering
\subfigure[]{\includegraphics[width=0.75\columnwidth,angle=-90]{Z9.6_R00B00.eps}\label{fig:GWa}}
\hfill
\subfigure[]{\includegraphics[width=0.75\columnwidth,angle=-90]{S11.2_R00B00.eps}\label{fig:GWb}}
\\
\subfigure[]{\includegraphics[width=0.75\columnwidth,angle=-90]{S50.0_R00B00.eps}\label{fig:GWc}}
\hfill
\subfigure[]{\includegraphics[width=0.75\columnwidth,angle=-90]{Z70.0_R00B00.eps}\label{fig:GWd}}
\caption{Gravitational waveform $A_+$ (top panels) and the corresponding spectrogram (bottom panels) for all models, labeled in each top-panel.
Here, we show only the non-vanishing component in axisymmetric profile $A_+\equiv Dh_+$ observed along the equatorial plane, which is perpendicular to the symmetric axis.
The spectrogram is obtained by short-time Fourier transform.
$D$ and $h_+$ are the source distance and gravitational wave strain, respectively, where the latter is calculated from a standard quadrupole formula.
For $s50$, we also plot a magnified view of the gravitational waveform with respect to the second bounce time in the inner mini-panel of the top-panel.
\label{fig:GW}}
\end{figure*}

\begin{figure*}[t!]
\centering
\includegraphics[width=0.7\columnwidth,angle=-90]{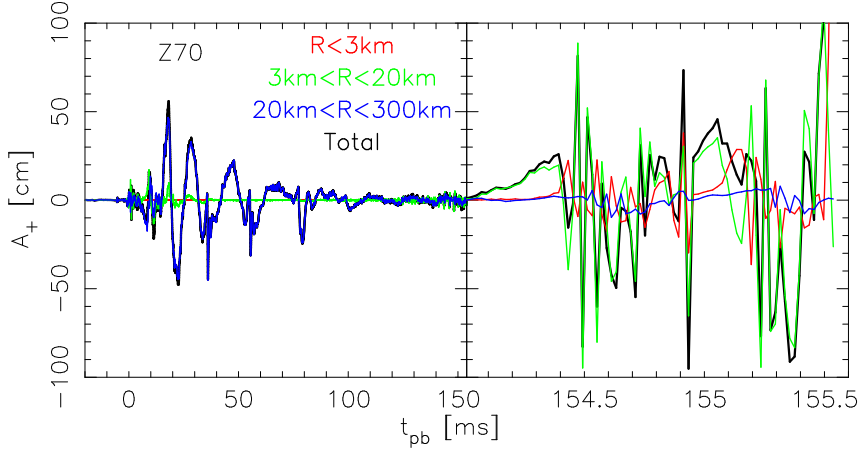}
\caption{GW amplitudes $A_+$ for the model $z70$ contributed from different spherical shells: innermost region with $r\leq3$\,km (red line), $3$\,km $<r\leq20$\,km (green line), and $20$\,km $<r\leq~300$\,km (blue line). Black line indicates the total gravitational waveform.
\label{fig:GW_Z70_atBH}}
\end{figure*}


\section{GW and neutrino emission}
\label{sec:GW and neutrino emission}
In this section we discuss the impact of the hadron-quark phase transition on the emission of gravitational waves and neutrinos.
In Fig.~\ref{fig:GW}, we plot gravitational waveforms and their spectrogram for all models.
Here, we show only a non-vanishing component $A_+=Dh_+$ observed along the equatorial plane that is perpendicular to the symmetric axis.
$D$ and $h_+$ are the source distance and gravitational wave strain, respectively.
$h_+$ is evaluated from a standard quadrupole formula \citep{Shibata&Sekiguchi03,KurodaT14}.

A common feature seen among all models is relatively large GW amplitudes, observed after the first core bounce till $t_{\rm pb}\sim50$\,ms.
These GWs are emitted from the postbounce convective motions and their amplitudes reach $\sim20-50$\,cm.
Afterward, the gravitational waveforms show various features depending on the progenitor mass.
Our least massive star $z9.6$, which is the NDE model and experiences the runaway explosion at around $t_{\rm pb}\sim100$\,ms (see, Fig.~\ref{fig:RshockEexp}), shows a considerable subsidence of GWs.
Using a similar progenitor mass model of $9$\,$M_\odot$, \cite{Vartanyan19b,Andresen21} also reported the subsidence of GWs. The authors found that it is due to a termination of mass accretion.
Furthermore the waveform is quite similar with that of a 2D relativistic model of \cite{BMuller13}.

Another NDE model $s11.2$ presents continuous GW emission even after the shock front exceeds $10^8$\,cm at $t_{\rm pb}\sim500$\,ms.
Furthermore, the waveform seems to show an increasing frequency, which can be indeed confirmed from the spectrogram in the lower panel.
The peak frequency concentrates around $\sim200$\,Hz at $t_{\rm pb}=200$\,ms and is increasing with time.
Such a ramp-up feature of the GW frequency is widely seen in CCSN models \citep{Murphy09,BMuller13,KurodaT16ApJL,Vartanyan19b,Mezzacappa20,Shibagaki21} and may come from the $g/f$-mode oscillation of the non-rotating PNS \citep{Morozova18,Torres-Forne19,Sotani20b} or perhaps PNS convection \citep{Mezzacappa20}. 
In the late post bounce phase $t_{\rm pb}\gtrsim600$\,ms, the spectrogram shows a wide band emission ranging from $300$\,Hz to a few kHz.
Since we use a simple short time Fourier analysis to obtain the spectrogram, we need to apply sophisticated analysis methods, such as S-method \citep{Kawahara18}, the Hilbert-Huang transform \citep{Takeda21}, or non-harmonic analysis \citep{Yanagisawa19}
to determine more detailed peak frequency.

Regarding the PDE model $s50$, the ramp-up feature can also be seen in the first $\sim200$\,ms after bounce.
It is remarkable that the ramp-up mode is showing a more rapid evolution in its peak frequency than that in $s11.2$.
At $t_{\rm pb}\sim180$\,ms, the peak frequency appears at $\sim500$\,Hz, which is significantly higher than $\sim200$\,Hz for $s11.2$.
The higher frequency in model $s50$ is simply due to a higher compactness of the PNS (see Table~\ref{tab:model_list}) and its higher $g/f$-mode oscillations.
From $t_{\rm pb}\sim200$\,ms to the second bounce time $t_{\rm pb}\sim380$\,ms, the GW emission shows a quiescent phase, during which the amplitudes are basically less than $\sim1$\,cm.
Such a quiescent phase can also be seen in our most massive star case $z70$ at $t_{\rm pb}\gtrsim100$\,ms.
In both models, the quiescent phase roughly corresponds to the period when the shock gradually recedes (Fig.~\ref{fig:RshockEexp}).
Although this paper does not aim to clarify the physical mechanism of the quiescent phase, we speculate that the rapid shock recession may suppress the convective motions, which usually excite the PNS core $g/f$-mode oscillations.
As a consequence, the weaker GWs are observed.
However, in the full 3D models, the situation may drastically change, as another degree of spatial freedom allows for the development of the standing accretion shock instability (SASI) \citep{Blondin03}.
In such models, the more compact PNS core is more susceptible to the SASI development.
The SASI activity can then imprint its feature in the emitted GW with sizeable amplitudes, typically in the low frequency range of $\sim100-200$\,Hz \citep{KurodaT16ApJL,Andresen17,Vartanyan19b,Mezzacappa20,Shibagaki21}.

The quiescent phase is disrupted by the hadron-quark phase transition.
Both models $s50$ and $z70$ experience the second collapse at $t_{\rm pb}\sim380$\,ms and $t_{\rm pb}\sim155$\,ms, respectively.
Although $z70$ immediately forms a BH, both models produce a quark core at their center and exhibit core bounce and subsequent ring down phase.
In inset of the upper panel for $s50$, we show a magnified view of the waveform at the second bounce.
The $x$-axis measures the time relative to the second bounce in millisecond.
The GW amplitudes reach $A_+\sim250$\,cm and the strong emission lasts about 4\,ms after the second bounce time.
Its burst like feature results in a wide band emission, ranging from $\sim500$\,Hz to beyond $2.5$\,kHz, where the latter value corresponds to the Nyquist frequency of our sampling rate of 5 kHz.
Such high frequencies are comparable with typical frequencies of the convection motions behind the second bounce shock as shown in Fig.~\ref{fig:BVfrequency}, where we see $|\omega_{\rm BV}|\sim$\,a few\,kHz at $6\times10^5$\,cm$\lesssim r\lesssim9\times10^5$\,cm.
Therefore these high-frequency GWs could be an outcome of these convection motions induced by the hadron-quark phase transition.
During the post second bounce phase in $s50$, the GWs show a high frequency emission at the range of $\gtrsim1$\,kHz and they subside after $t_{\rm pb(p2b)}\sim460(80)$\,ms.
The maximum amplitude and the duration time of the strong emission are comparable to those in \cite{Zha20}.

Regarding the BH formation model $z70$, it also shows a high peak frequency $\gtrsim1$\,kHz and their amplitudes reach $A_+\sim100$\,cm.
In contrast to the full 3D BH formation models \citep{KurodaT18,Shibagaki21}, whose GW waveforms are contaminated by the low-frequency SASI modes, the GW signal at the BH formation shows a higher peak frequency \citep[this could be consistent with][]{sotani19,sotani21}.
This is not only due to the absence of significant SASI motions but also to the quark core formation and its oscillation.
We crudely check if the GW emission prior to the BH formation actually emanate from outside the apparent horizon.
In Fig.~\ref{fig:GW_Z70_atBH}, we show the GW amplitudes $A_+$ for model $z70$ contributed from different spherical shells: innermost region with $r\leq3$\,km (red line), $3$\,km $<r\leq20$\,km (green), and $20$\,km $<r\leq300$\,km (blue).
Black line indicates the total gravitational waveform.
To plot the figure, we simply divide the computational domain into several spherical shells, calculate the contained mass quadrupole moment inside each shell, and derive the gravitational wave strain.

One can see that the blue line overlaps with the black one in most of the post bounce phase up to $t_{\rm pb}=150$\,ms.
It indicates that the dominant contribution to the total GW is coming from the post shocked region, roughly corresponding to the region with $20$\,km $<r\leq300$\,km.
While the inner region with $r\leq20$\,km shows essentially no GW emission except during the initial post convection phase ($t_{\rm pb}\lesssim30$\,ms).
However, just prior to the second collapse ($t_{\rm pb}\geq154$\,ms, i.e. less than $\sim1$\,ms till the BH formation), the dominant contribution is from $r\leq20$\,km.
At this moment, the quark core is already formed.
Although the contribution from the innermost region with $r\leq3$\,km is not negligible (see the red line), the majority comes from outside of the apparent horizon, i.e. $3$\,km  $<r\leq20$\,km.
Here we note that the circumferential radius of AH is $\sim2.29$\,km at the final simulation time.
Therefore, the signal of the QCD phase transition could be imprinted into the emanated GWs, even in the failed CCSN models.
We, however, stress that the gravitational waveform can greatly change in the full 3D models and also that the gravitational wave should be evaluated more precisely by the gauge invariant method, especially in such a model on a highly curved spacetime.

\begin{figure}[htbp]
\begin{center}
\includegraphics[width=70mm,angle=-90.]{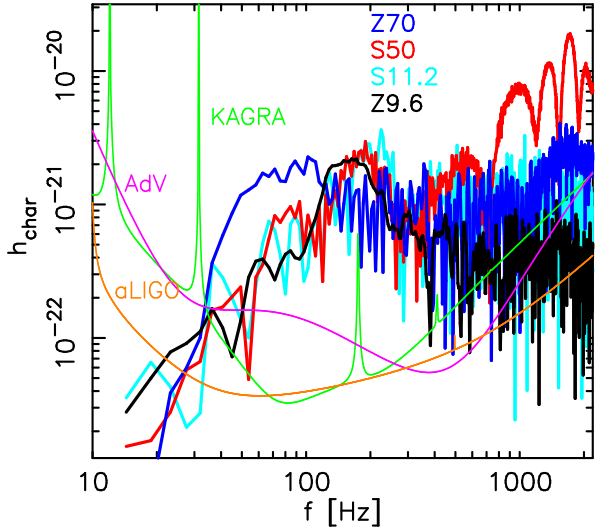}
\caption{Characteristic GW spectral amplitudes for all models assuming a source distance of 10\,kpc.
The noise amplitudes of aLIGO (orange line), AdV (magenta line), and KAGRA (green line) are plotted as reference.
Note that, since the Nyquist frequency is 2500 Hz, we show only $f\lesssim2000$\,Hz.
\label{fig:hchar}}
\end{center}
\end{figure}

To discuss the detectability of the GW signals, we show the GW spectral amplitudes for all models in Fig.~\ref{fig:hchar}, assuming a source distance of 10\,kpc, together with the sensitivity curves of the advanced LIGO (aLIGO), advanced VIRGO (AdV), and KAGRA \citep{abbott18det}.
The common feature seen among the models is a spectral peak appearing in a relatively low frequency range of $\sim100-200$\,Hz.
From the spectrogram in Fig.~\ref{fig:GW}, such low frequency GWs are emitted mainly by convection motions in the postbounce phase.
Our most massive progenitor model $z70$ shows a relatively lower peak frequency compared to the other models appearing at $50\lesssim f\lesssim100$\,Hz.
This can be explained by the less compact PNS for model $z70$, although the PNS soon becomes more compact than the rest of models due to a higher mass accretion rate.

Once the initial convection motion ceases, the GW frequency increases with time.
While $z9.6$ has only one spectral peak at $f\sim150$\,Hz, due to a considerable subsidence of GWs after $t_{\rm pb}\gtrsim100$\,ms, the other models exhibit a broadband emission ($s11.2$), extending from $\sim200$\,Hz to $\gtrsim1000$\,Hz, or rather a double-peaked feature ($s50$ and $z70$).
The high frequency emission seen in model $s11.2$ has nearly a comparable amplitude $h_{\rm char}\sim2\times10^{-21}$ with that of the low frequency emission.
It is emitted in the late post bounce phase, $t_{\rm pb}\gtrsim300$\,ms from Fig.~\ref{fig:GWb}, and mainly originated from the oscillation of contracting PNS or its convection, which is still under debate.
$s50$ and $z70$ also have a high frequency spectral peak appearing at $\sim1700$\,Hz.
From Fig.~\ref{fig:GW}, such a high frequency emission occurs during the second collapse and is associated with the hadron-quark phase transition and its subsequent convection motions as already explained (see also Fig.~\ref{fig:BVfrequency}).
The amplitude at $f\gtrsim1000$\,Hz in model $s50$ is well above the sensitivity curves of the current-generation GW detectors and could be detectable, if the source locates within our Galaxy.
The present GW analysis for $s50$ is in qualitative agreement with the results reported by \citet{Zha20}, in particular the location of the secondary peaks in excess of 1000\,Hz which are related with the dynamics of hadron-quark phase transition.

\begin{figure*}[t!]
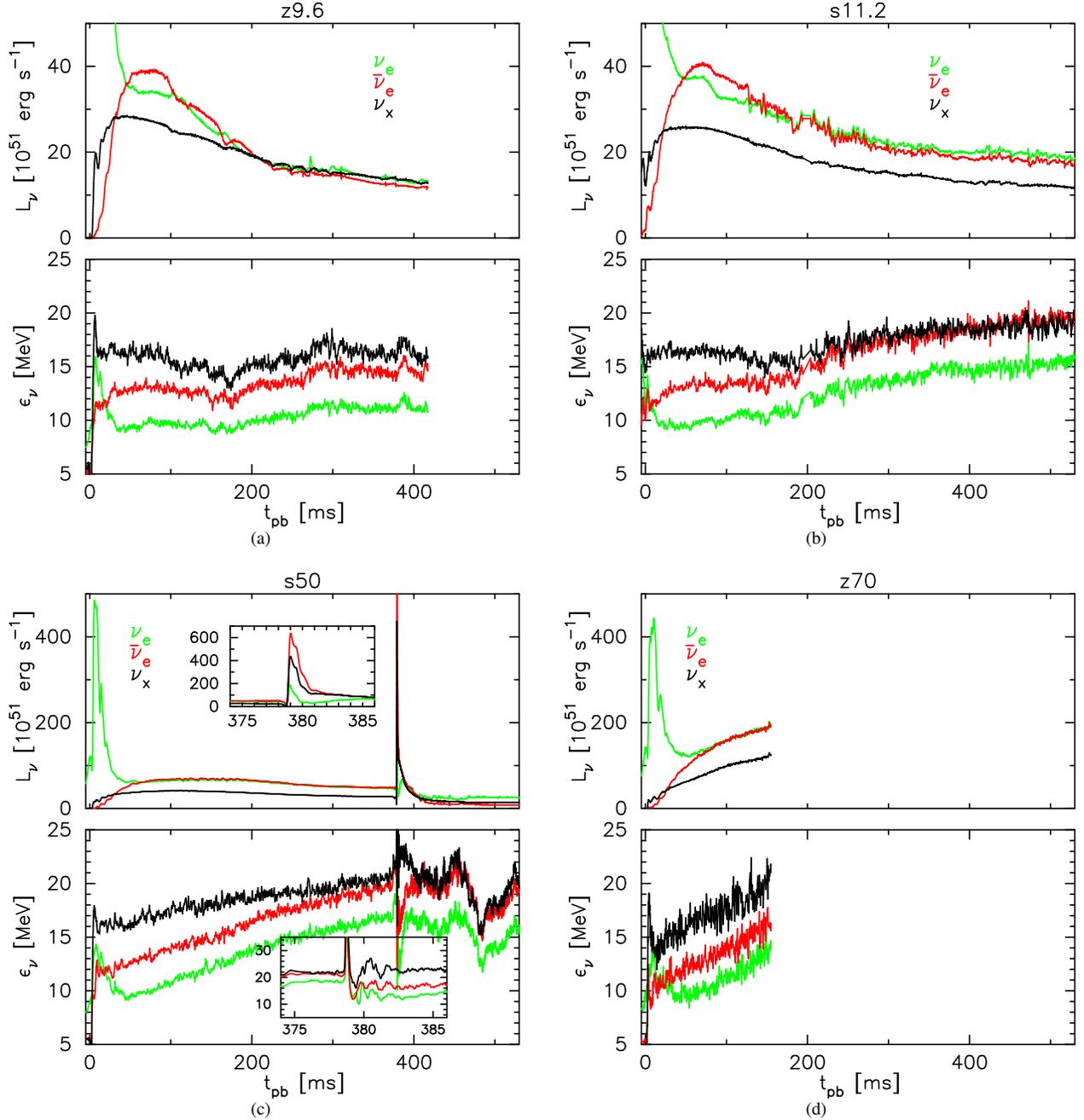

\centering
\subfigure[]{\includegraphics[angle=-90.,width=80mm]{LnuEnuz9.6.eps}\label{fig:LnuEnu_z96}}
\hspace{5mm}
\subfigure[]{\includegraphics[angle=-90.,width=80mm]{LnuEnus11.2.eps}\label{fig:LnuEnu_s11.2}}
\\
\subfigure[]{\includegraphics[angle=-90.,width=80mm]{LnuEnus50.eps}\label{fig:LnuEnu_s50}}
\hspace{5mm}
\subfigure[]{\includegraphics[angle=-90.,width=80mm]{LnuEnuz70.eps}\label{fig:LnuEnu_z70}}
\caption{Post-bounce evolution of the neutrino luminosities (top panels) and mean neutrino energies (bottom panels) for all models. Green, red, and black lines belong to $\nu_e$, $\bar\nu_e$, and $\nu_x$, respectively.
For $s50$, the inlays show the second burst.
\label{fig:LnuEnu}
}
\end{figure*}

Fig.~\ref{fig:LnuEnu} presents the neutrino luminosities and mean energies for all models\footnote{We have checked the reliability of current neutrino transport method in the Cartoon coordinates using the standard neutrino opacity set, which is summarized in App.~\ref{app:Neutrino Check}.}.
All quantities are evaluated at a distance of $r=400$\,km.
Comparing with full GR 2D models in \cite{BMuller14}, the less massive progenitor models $z9.6$ and $s11.2$ show very similar neutrino profiles.
We note that the detailed neutrino opacities are different between this study and theirs.
The neutrino luminosities of all flavors overlap each other at $t_{\rm pb}\gtrsim200$\,ms in model $z9.6$.
The neutrino mean energies show only a slight increase with time, which indicates that the PNS core does not shrink significantly in this model.
At $t_{\rm pb}=400$\,ms, $L_{\nu_e,\bar\nu_e,\nu_x}$ become $\sim1.2\times10^{52}$\,erg\,s$^{-1}$ and mean energies reach $\varepsilon_\nu\sim11$, 14, and 16 MeV for $\nu_e$, $\bar\nu_e$, and $\nu_x$, respectively.
In $s11.2$, the accretion luminosities peak at around $t_{\rm pb}\sim90$ and reach $L_{\nu_e,\bar\nu_e}\sim4\times10^{52}$\,erg\,s$^{-1}$ and $L_{\nu_x}\sim2.5\times10^{52}$\,erg\,s$^{-1}$.
Afterward they show a decreasing trend.
The mean neutrino energies become higher with time and $\varepsilon_{\bar\nu_e}$ reaches nearly the same value with $\varepsilon_{\nu_x}$ at $t_{\rm pb}\gtrsim300$\,ms.
All of these features seen in both $z9.6$ and $s11.2$ are roughly in good agreement with those in \cite{BMuller14}.

$s50$ shows a quite remarkable feature in association with the hadron-quark phase transition.
After the first bounce, $L_{\nu_e}$ and $L_{\bar\nu_e}$ become almost the same values at $t_{\rm pb}\gtrsim60$\,ms, while $L_{\nu_x}$ shows a value of nearly half of the electron type (anti-)neutrino luminosities.
When the second collapse initiates, the neutrino luminosities of all flavors drastically increase.
From inset panel, we see that $L_{\bar\nu_e}$ and $L_{\nu_e}$ are showing the highest and lowest luminosity, respectively, and such a hierarchy is consistent with \cite{Fischer20}.
The hierarchy can be explained by the neutron rich environment, through which the shock propagates.
As we have explained in Fig.\ref{fig:QcbM50bb}, the shock is generated beneath the $Y_e$-trough and propagates through the neutron rich environment.
Therefore abundant electron type antineutrinos are produced via the inverse beta decay process, leading to the highest $L_{\bar\nu_e}$.
The peak luminosity of $L_{\bar\nu_e}\sim6\times10^{53}$\,erg\,s$^{-1}$ at the second bounce is nearly the same order of that of $L_{\nu_e}\sim5\times10^{53}$\,erg\,s$^{-1}$ at the first bounce.
Such a feature is also reported in \cite{Fischer20}.

The mean neutrino energies in our most compact model $z70$ show the fastest increase.
This is due to the most rapid PNS contraction among the models and to a hottest PNS core as a consequence \citep{Liebendorfer04,Sumiyoshi07}.
\cite{Shibagaki21} reported the neutrino luminosities for the same $70$\,M$_\odot$ progenitor model with different nuclear EOS and neutrino opacities.
Compared to their results, the neutrino luminosities show slightly higher values of $\sim20$\,\% in all flavors.
The neutrino signal associated with the QCD phase transition is not plotted in the figure.
This is because we evaluate the neutrino flux at $r=400$\,km and the simulation stops before the neutrino burst reaches that sphere.

\begin{figure}[htbp]
\begin{center}
\includegraphics[width=80mm,angle=-90.]{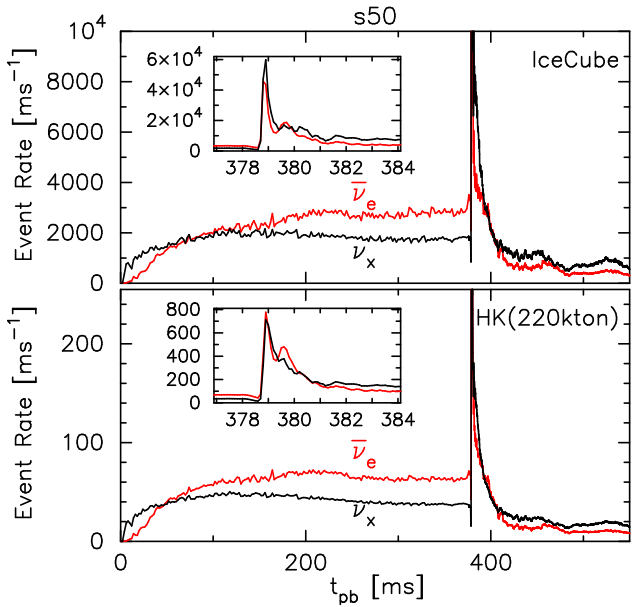}
\caption{Neutrino detection rate of IceCube (top panel) and HK (bottom panel) as a function of time after bounce for model $s50$ observed at 10\,kpc.
The red and black lines are for $\bar\nu_e$, assuming no flavor conversion, and for $\bar\nu_x$, assuming complete flavor conversion, respectively.
The inlays show the second burst.
\label{fig:CountRate}}
\end{center}
\end{figure}
Finally we discuss the detectability of the second neutrino burst in association with the hadron-quark phase transition for model $s50$.
Fig.~\ref{fig:CountRate} illustrates the neutrino detection rate of IceCube (top panel) and HK (bottom panel) as a function of post bounce time for model $s50$ observed at 10\,kpc.
We evaluate the event rates of each model for two neutrino detectors: IceCube \citep{abbasi11,salathe12} and Hyper-Kamiokande (\citet{abe11,HK18}, hereafter HK).
The main channel in these detectors is of anti-electron neutrino ($\bar\nu_e$) with inverse-beta decay.
For simplicity, we take into account only this channel to evaluate the event rates.
We assume that the neutrino energy spectrum is a Fermi-Dirac distribution (see \citet{lund10,takiwaki18} for more detail).
Furthermore in the evaluation, we consider two extreme cases: all $\bar\nu_e$ emitted from the source reach the detectors without neutrino flavor conversion and cause the signal at the detectors (red lines in the figure); all $\bar\nu_x$ (identical to $\nu_x$ in this study) emitted from the source are completely swapped by $\bar\nu_e$ and cause the signals (black lines).

From the plot, we find that, during the first $\sim50$\,ms after bounce, the $\bar\nu_x$ signal shows a larger detection rate for both detectors.
Afterward the $\bar\nu_e$ signal becomes stronger than that of $\bar\nu_x$.
After $t_{\rm pb}\sim100$\,ms, both signals show nearly constant values until the onset of the second collapse.
Such features are consistent with previous normal CCSN models \citep{Tamborra13,Nagakura21_neutrino_signal}.
Once the second collapse takes place, the signals become approximately 30 times stronger than the values before the second collapse.
The peak values for both $\bar\nu_e$ and $\bar\nu_x$ signals reach nearly the same value of $\sim5000$\,ms$^{-1}$ (IceCube) and $\sim700$\,ms$^{-1}$ (HK).
\cite{Fischer18} reported the number of expected neutrino events in the Super-Kamiokande (SK) detector of $\sim100$ with 2\,ms bins from the second neutrino burst event, assuming a source distance of 10 kpc.
Concerning the fiducial volume of HK, which is approximately 10 times larger than that of SK, our expected event rate $\sim700$\,ms$^{-1}$ in HK can be converted to $\sim140$ signals with 2\,ms bins, which is in line with that of \cite{Fischer18}.

\section{Summary and Discussion}
\label{sec:Summary and Discussion}

In this study, we have presented possible effects of the phase transition from hadronic matter to quark matter in CCSNe. For this purpose, we have performed 2D axisymmetric numerical simulations of core collapse of massive stars with $9.6$\,M$_\odot$, $11.2$\,M$_\odot$, $50$\,M$_\odot$, and $70$\,M$_\odot$. Our core-collapse supernova model is based on numerical relativity, which solves the general relativistic neutrino radiation hydrodynamics equations together with the two-moment (M1) neutrino transport equations of \citet{KurodaT16}. We use up-to-date neutrino opacities following \cite{Kotake18}. We employed the  hybrid EOS DD2F--RDF~1.2 of \citet{Fischer18} and \citet{Bastian:2021}, which considers a first order hadron-quark phase transition.

Our results demonstrate a diversity of post bounce evolution depending on the progenitor.
We confirm that low and intermediate mass progenitors, examined at the examples of a low-metalicity model with ZAMS mass of 9.6\,M$_\odot$ and a solar-metalicity model with ZAMS mass of 11.2\,M$_\odot$, belong to the class of core-collapse supernova explosion driven by the neutrino-heating mechanism. The latter one is aided by the presence and development of convection. Both neutrino-driven explosion models are in qualitative agreement with previous studies, e.g., in terms of the explosion dynamics as well as the diagnostic explosion energy obtained.
None of these models reach the conditions required for the hadron-quark phase transition, prior to the onset of the supernova explosion. In contract, the more massive progenitors explored with ZAMS mass of 50\,M$_\odot$ of solar metalicity and an extremely metal-poor star with ZAMS mass of 70\,M$_\odot$, both undergo the phase transition during the post-bounce mass accretion phase. However, an explosion is only obtained for the 50\,M$_\odot$ model while for the 70\,M$_\odot$ model we observe a failed explosion and black-hole formation instead. This is related to the maximum mass of the DD2F--RDF~1.2 hybrid EOS, which the latter one exceeds at the onset of the hadron-quark phase transition. Contrary therefore, for the 50\,M$_\odot$ model we report the formation and propagation of a second shock wave, as a direct consequence of the hardon-quark phase transition, which determines the onset of the supernova explosion, in qualitative agreement with the results reported by \citet{Fischer18} and \citet{Fischer:2020}. Nevertheless, the yet incompletely understood question about the dependence on metalicity, preliminarly addressed in \citet{Fischer:2021}, remains to be explored in a more systematic study.



Our finding inherent to the multi-D nature is the strong RT mixing.
In the context of CCSNe, the potential region of development of the RT mixing is usually distant from the PNS, e.g. the interface between CO core and He layer \citep{Wongwathanarat15}, which locates at $r\gtrsim10^{9-10}$\,cm depending on the progenitor model.
However, because of the shallow density profile in $s50$, the RT mixing can develop even at a distance of $r\sim10^8$\,cm.
Because of the short distance, the RT mixing can develop in the early post explosion phase of a few 10\,ms after the explosion onset has been launched in terms of the initial propagation of the second shock wave.
A combination of the PDE and a prompt development of the RT mixing might have an influence on the nucleosynthetic yields previously reported in \citet{Fischer:2020}.

Furthermore, we discussed possible effects of the  hadron-quark phase transition on the multi-messenger observables, gravitational waves and neutrinos. For the less massive progenitor models $s9.6$ and $s11.2$, we confirm commonly found features in their GW signals: $z9.6$ shows a subsidence of GW emission, once it entered a runaway explosion phase at $t_{\rm pb}\gtrsim200$\,ms, and $s11.2$ exhibited a rather constant GW emission even after its explosion, though the peak frequency increases with time in association with the PNS contraction.
The massive progenitor stars $s50$ and $z70$ display a quiescent phase for GW emission once their standing shock front starts to recede. However, in full 3D models, the SASI phase could arise in this period and contribute to the GW emission. The contribution can be significant to produce sizeable GWs according to the previous 3D studies \citep[e.g.,][]{KurodaT16ApJL}. At the second bounce in $s50$, ring down of the quark core produces strong and high-frequency GWs, with their amplitudes and typical frequencies exceeding $Dh\sim250$\,cm and 1\,kHz, respectively.
Since the Brunt-V\"ais\"al\"a frequencies behind the second bounce shock are reaching a few\,kHz, the high-frequency GWs could be an outcome of these convection motions \cite[see also the supplemental material of][for the relation between the convection motions behind the second bounce shock and typical GW frequencies.]{Zha20}.
Although their typical frequency appears at kHz range, which is out of the best sensitivity of current ground-based gravitational wave detectors, their signal is strong enough to be detected, if the source locates at 10 kpc.
Even in the failed explosion case $z70$, the region outside the apparent horizon contributes significant GW emission.
Thus, we can expect that the signal of a hadron-qurk phase transition might be imprinted in GWs even in the failed CCSNe, although the detailed waveform should be examined in full 3D models, considering the SASI development.

Concerning the neutrino profiles, the neutrino luminosity and mean energy of all flavors have shown quite reasonable values in models $z9.6$ and $s11.2$ with previous studies \citep[e.g. with 2D relativistic models of][]{BMuller14}, since these two less massive progenitor models do not undergo the hadron-quark phase transition during our simulation time.
While in $s50$, we have observed the second neutrino burst when the second core-bounce takes place.
In contrast to the first bounce, the electron type antineutrino luminosity $L_{\bar\nu_e}$ exhibits the highest luminosity among the flavors.
This is because the quark core bounce occurs below the $Y_e$-trough, and the shock propagates through the neutron rich environment.
Therefore, the inverse beta decay becomes the dominant reaction process and produces numerous electron type antineutrinos.
The hierarchy seen in the second neutrino burst is consistent with previous studies \citep{Fischer18,Zha20}.

Recently long-time 2D SN simulations, longer than several seconds in 2D \citep[c.f.][]{BMuller15b,Nakamura19,Burrows21,Nagakura21_longterm2D} and even in 3D by \citet{Bollig21_7min} become computationally feasible.
Because of our short simulation times, the diagnostic explosion energy of models $z9.6$ and $s11.2$ do not reach energy saturation.
According to \cite{BMuller15b}, the energy saturation can be reached after $\sim$5~s for the model $s11.2$.
Such long-time simulations are necessary to discuss the diagnostic explosion energy, as well as the final property of remnant star. 
It is worth mentioning that the inclusion of progenitor's precollapse inhomogenities in the burning shells \citep{Yadav20,Yoshida19,Yoshida20, Yoshida21} has been pointed out to make the underpowered explosion more energetic \citep{bernhard14,Bollig21_7min}.
Starting from the multi-D progenitors, we have to study how the multi-D hydrodynamic motions in the pre-explosion phase result in the  multi-messenger signals up to the late explosion phase (see previous attempts in \citet{Nakamura19}, \citet{Nagakura21_longterm2D}, and  \citet{MaxW21}). Though this is not an easy task, this should be the only promising way we should be on to enhance the predictive power of CCSN multi-messenger signals as quantitatively as possible.

\acknowledgements{
We thank the referee Shuai Zha for clarifications that improved this article.
We are also grateful to N.U. Bastian for providing (extending) the EOS table, which  is readily employed in the multi-D CCSN simulations.
T.K., T.T., and K.K. are thankful to H. Ookawa for helpful discussions in developing the apparent horizon finder in this work.
Numerical computations were carried out on Cray XC50 at CfCA, NAOJ and also on Sakura and Raven at Max Planck Computing and Data Facility.
T.F. acknowledges support from the Polish National Science Center (NCN) under Grant No. 2019/33/B/ST9/03059.
This work was supported in part by Japan Society for the Promotion of Science (JSPS) KAKENHI Grant Numbers JP17H06357, 
JP17H06364, 
JP18H01212, 
JP21H01088. 
K.K. was supported by Research Institute of
Stellar Explosive Phenomena at Fukuoka University and the associated project (No.207002).
}

\appendix
\section{Spatial resolution test}
\label{app:Spatial resolution dependence}
In this appendix, we are showing spatial resolution dependence of the post-bounce PNS evolution to explain that the currently used spatial resolution is adequate to follow the quark core evolution.
For that purpose, we perform four CCSN simulations ``$L09$--$12$'' with different spatial resolution.
The integer after $L$ represents the maximum refinement level set for the nested structure, e.g. $L10$ corresponds to a model with the maximum refinement level of $L_{\rm max}=10$.
The central grid width is halved by raising the level by one.
As a result, the central numerical grid width achieves $\sim 460$, $230$, $115$, and $57$\,m for $L09$, $L10$, $L11$, and $L12$, respectively.
We, however, do not run these models with a static nested structure from initial, instead we raise the level of nested structure depending on the value of maximum rest mass density as follows.
In $L09$ and $L10$, we do not change the maximum refinement level throughout the calculation.
These models are calculated to see the resolution dependence mainly on the PNS contraction phase including the second collapse time.
In $L11$ and $L12$, we restart $L10$ shortly before its maximum rest mass density $\rho_{\rm max}$ reaches $8\times10^{14}$\,g\,cm$^{-3}$.
Then we raise the level to $L_{\rm max}=11$ once $\rho_{\rm max}$ exceeds $8\times10^{14}$\,g\,cm$^{-3}$.
Furthermore in $L12$, we again raise the level to $L_{\rm max}=12$ when $\rho_{\rm max}$ exceeds $9\times10^{14}$\,g\,cm$^{-3}$.
$L11$ is identical to the model $s50$ discussed in the main text and $L10$--$12$ are calculated to check the resolution dependence particularly focusing on the quark core formation and subsequent evolution phase.

\begin{figure*}[t!]
\begin{center}
\includegraphics[width=0.35\columnwidth,angle=-90.]{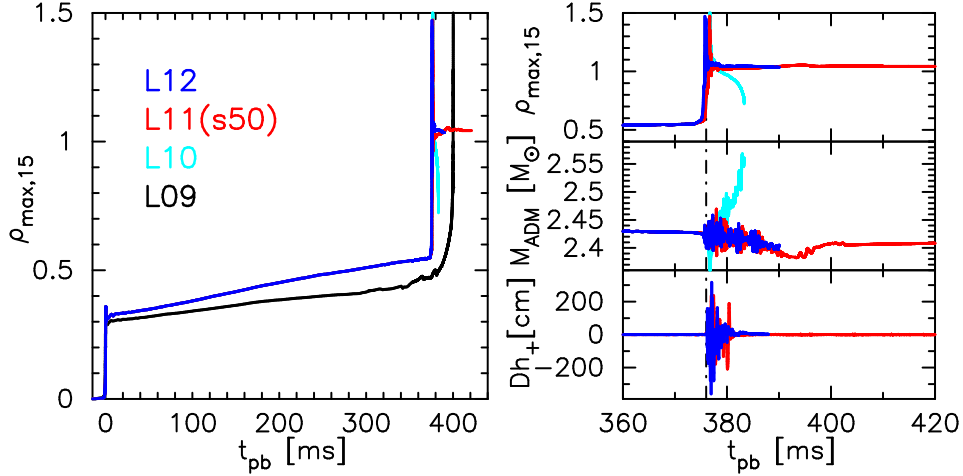}
\caption{{\em Left panel}:~Evolution of the maximum density $\rho_{\rm max,15}(\equiv \rho_{\rm max}/10^{15})$ for four cases $L09$--$L12$ (see text for their numerical setup). {\em Right panel}: Zoom in for the evolution of the maximum density $\rho_{\rm max,15}$ (top panel) and ADM mass M$_{\rm ADM}$ (bottom panel) for $L12$ (blue line), $L11$ (red line), and $L10$ (light blue), focusing on the post second bounce phase. The vertical dot-dashed lines indicate when we raise the maximum refinement level from $L_{\rm max}=10$ to 11.
\label{fig:App1}
}
\end{center}
\end{figure*}

In the left panel in Fig.~\ref{fig:App1}, the maximum density evolution is plotted for four cases $L09$--$L12$.
In the right panel, we show the maximum density (top), ADM mass (middle), and GW amplitudes $A_+$ (bottom) for $L12$ (blue line), $L11$ (red line), and $L10$ (light blue), focusing on the post second bounce phase.
As for the gravitational waveform, we plot only for $L11$ and $L12$.
We first check the impact of numerical resolution on the PNS contraction phase and also on the second collapse time.
Since there is essentially no difference in the second collapse time among the three models $L10$--$L12$, here we compare $L12$ (blue) and $L10$ (black) in the left panel.
The finest numerical grid in these two models is $\sim230$\,m (blue) or $\sim450$\,m (black) during the PNS contraction phase ($t_{\rm pb}\lesssim350$\,ms).
From the left panel, the black line systematically shows a smaller maximum density compared to that of $L12$ (note that the blue, red, and light blue liens are identical when $\rho_{\rm max,15}<0.8$).
Although the difference already appears at just after the first bounce, we also find that the central entropy secularly increases in $L09$ (not shown in the figure).
This is due to the numerical dissipation in lower resolution model $L09$.
Therefore, the maximum density in $L09$ is showing smaller value throughout the PNS contraction phase.
However the second collapse times in $L10$--$12$ and $L09$ are exhibiting a rough convergence, only $\sim 20$\,ms difference between red and black lines.
We can thus say that, for the PNS contraction phase, a rough numerical convergence is achieved for grid width smaller than $\sim 450$\,m.

Next we compare $L10$, $L11$, and $L12$ focusing on the numerical convergence after the second core bounce.
From top right panel, it is obvious that the maximum density drastically decreases after the second core bounce in $L10$ (light blue).
At the same time, the ADM mass significantly increases and does not settle into a new quark core branch as $L11$ does.
This is because the lower resolution model $L10$ suffers from a severe numerical dissipation and the quark core seems to be artificially gravitationally unbound.
From a comparison between $L11$ and $L12$, though we cannot follow a long time evolution for the highest numerical resolution model $L12$, we apparently see a convergence in the post second bounce evolution at least during the first $\sim15$\,ms.
Both the maximum rest mass density and the ADM mass are exhibiting consistent evolutions.
In addition from bottom-right panel, the strong GW emissions associated with the second bounce are roughly converged among these two models. Indeed, the observed maximum amplitude $|A_+|\sim300$\,cm and duration time of $\sim5$\,ms are comparable in $L11$ and $L12$.
Since the emission considered here is due to stochastic motions during the ring down phase, a convergence in the maximum amplitude and duration time is sufficient for the spatial resolution check.
From these comparisons, we conclude that a resolution of $\sim110$\,m used in $L11(s50)$ is required to adequately resolve the quark core evolution.

\section{Test for the neutrino transport in a mixed frame of cylindrical and Cartoon coordinates}
\label{app:Neutrino Check}
\begin{figure*}[t!]
\begin{center}
\includegraphics[angle=-90.,width=0.95\columnwidth]{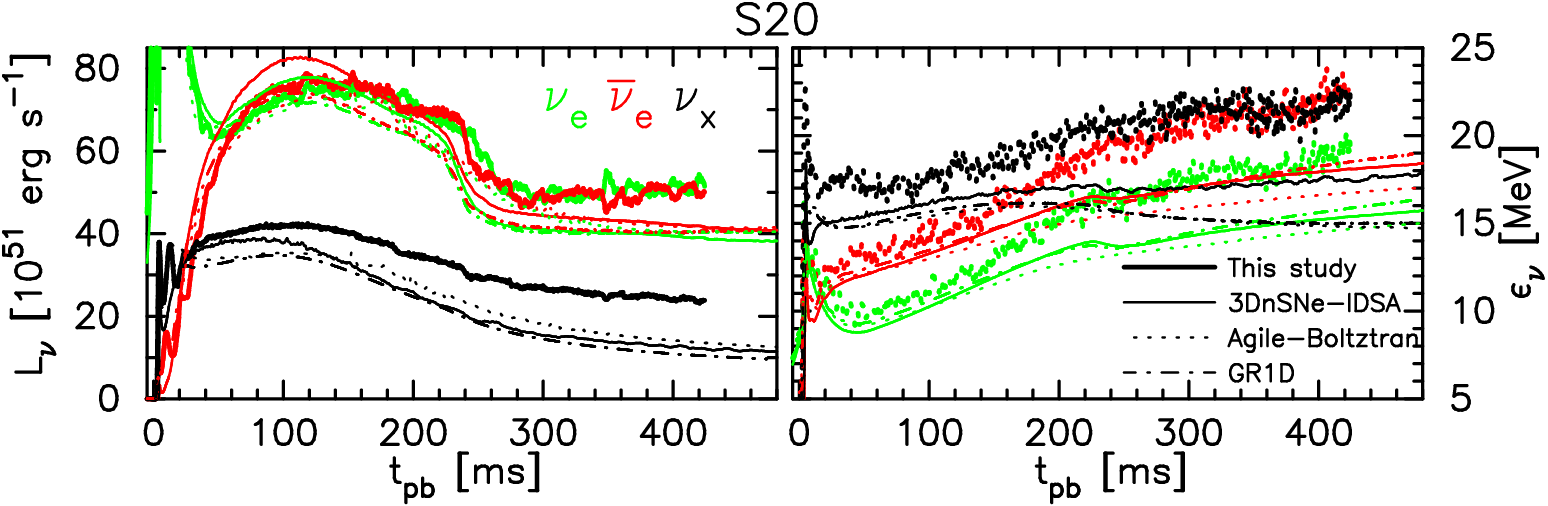}
\caption{Post-bounce evolution of the neutrino luminosities (left panel) and mean neutrino energies (right panel) for the solar metallicity, non-rotating $20$\,M$_\odot$ progenitor star from \citet{WH07}.
The latter is identical to the one used in a global comparison of CCSNe simulations reported by \citet{O'Connor18_GlobalComparison}.
The neutrino opacity is the same as was used in \citet{O'Connor18_GlobalComparison}. Green, red, and black lines belong to $\nu_e$, $\bar\nu_e$, and $\nu_x$, respectively.
Thick lines are results of this study, while thin, dotted, and dash-dotted lines are results using {\small 3DnSNe-IDSA} code \citep{takiwaki16}, {\small Agile-Boltztran} code \citep{Liebendorfer04}, and {\small GR1D} code \citep{O'Connor15_GR1D}, respectively.
\label{fig:LnuEnuS20}
}
\end{center}
\end{figure*}

This appendix is devoted to test our neutrino transport in a mixed frame of cylindrical and Cartoon coordinates, in which all hydrodynamical and neutrino-radiation quantities, as well as the metric variables, are not evolved straightforwardly as in the Cartesian coordinates used in our previous studies \citep[e.g.,][]{KurodaT16}.
To this end, we perform the same CCSN simulation reported by \cite{O'Connor18_GlobalComparison} and follow the collapse of the solar metallicity, non-rotating $20$\,M$_\odot$ star taken from \cite{WH07}, and compare the neutrino profiles.
Furthermore, the neutrino opacity is also the same with \cite{O'Connor18_GlobalComparison} and uses the baseline set of \cite{Bruenn85} including Brehmsstrahlung of \cite{hannestad98}.
The energy bins are 20 bins, which logarithmically cover the region of 0-300 MeV.
One of the major difference from \cite{O'Connor18_GlobalComparison} is the spherical symmetric property, which we cannot take into account in the Cartoon coordinates based numerical code.
We also note that the presented model is calculated by numerical relativity and thus might have some inherent differences from those in \cite{O'Connor18_GlobalComparison}.
Therefore, in the late post bounce phase, when the convection motions become dominant, our neutrino profiles show different results from those in \cite{O'Connor18_GlobalComparison}.
In Fig.~\ref{fig:LnuEnuS20}, we plot neutrino luminosities (left) and mean neutrino energies (right) extracted at a radius of $r=400$\,km by thick lines In addition, we plot three reference cases obtained by using following codes: {\small 3DnSNe-IDSA} \citep{takiwaki16} (thin line), {\small Agile-Boltztran} \citep{Liebendorfer04} (dotted line), and {\small GR1D} \citep{O'Connor15_GR1D} (dash-dotted line).
The data for these reference cases is downloaded from the publisher site of \cite{O'Connor18_GlobalComparison}.

We first explain comparison of neutrino luminosities.
From left panel, the accretion luminosity reaches $\sim7.5\times10^{52}$\,erg\,s$^{-1}$ for $\nu_e$ and $\bar\nu_e$ and $\sim4.1\times10^{52}$\,erg\,s$^{-1}$ for $\nu_x$.
The post bounce time, when those accretion luminosity reach their maximum, as well as their peak values are in good agreement with \cite{O'Connor18_GlobalComparison}.
At around $t_{\rm pb}\sim240$\,ms, the Si-O interface accretes and the mass accretion rate suddenly drops.
Consequently $L_{\nu_e,\bar\nu_e}$ significantly decreases.
Afterward it enters the convection dominant phase and the neutrino profile in multi-D model starts to diverge from that in spherically symmetric model.
Indeed due to the vigorous convection motions, our neutrino luminosites show systematically higher values in all flavors than those in 1D models of \cite{O'Connor18_GlobalComparison}, roughly $25$\,\% and $100$\,\% increase for $L_{\nu_e,\bar\nu_e}$ and $L_{\nu_x}$, respectively, at $t_{\rm pb}=400$\,ms.
Such a higher neutrino luminosity in multi-D model reflects the effects of PNS convection around the neutrino sphere \citep{Buras06a}.

Regarding the neutrino mean energy (right panel), we again find a consistent result with two reference models \cite{O'Connor18_GlobalComparison}, at least for the first $\sim100$\,ms after bounce.
During that early epoch, $\varepsilon_{\nu_e}$ reaches its minimum value of $\sim9$\,MeV at $t_{\rm pb}\sim50$\,ms.
$\varepsilon_{\bar\nu_e}$ is showing a higher value than $\varepsilon_{\nu_e}$, preserving the deviation $\varepsilon_{\bar\nu_e}-\varepsilon_{\nu_e}$ to be $\sim3$\,MeV throughout most of the post bounce phase.
This feature is also seen in \cite{O'Connor18_GlobalComparison}.
While $\varepsilon_{\nu_x}$ stays at $\sim17$\,MeV just after bounce $t_{\rm pb}\lesssim100$\,ms, and then gradually increases, though the increase rate is not so high as those of $\varepsilon_{\nu_e}$ and $\varepsilon_{\bar\nu_e}$.
Due to the multi-D effects, all the mean neutrino energies become systematically higher than the 1D results, especially in the late post bounce phase $t_{\rm pb}\gtrsim200$\,ms.
From these comparisons, we conclude that the new neutrino transport method in a mixed frame of the cylindrical and Cartoon coordinates works consistently with other numerical codes and is capable to perform CCSN simulations.

\bibliographystyle{aasjournal}
\bibliography{mybib}

\end{document}